\documentclass[preprintnumbers,prd,twocolumn,aps,superscriptaddress,nofootinbib,amsfonts,amsmath,amssymb,bm,showpacs,floatfix]{revtex4-1}

\usepackage{subfigure}
\usepackage{color,amsmath,amssymb,amsfonts,amsbsy}
\usepackage{bbm}
\usepackage{bm}
\usepackage[dvipsnames]{xcolor}
\usepackage[final]{graphics}
\usepackage{epsfig}
\usepackage{sidecap}
\usepackage[colorlinks=true,linkcolor=blue,citecolor=blue,urlcolor=blue]{hyperref}
\usepackage{lipsum}
\usepackage{graphicx} 
\usepackage{color}
\usepackage{subfigure}
\usepackage{relsize}
\usepackage{listings}
\usepackage{tabularx}
\usepackage{listings}
\usepackage{inconsolata}
\usepackage{dcolumn}
\usepackage{placeins} 

\makeatletter
\renewcommand*{\@fnsymbol}[1]{\ensuremath{\ifcase#1\or *\or \dagger\or \ddagger\or
   \mathsection\or \mathparagraph\or \|\or **\or \dagger\dagger 
   \or \ddagger\ddagger 
   \or \mathsection\mathsection
   \or \mathparagraph\mathparagraph
   \or \|\|
   \else\@ctrerr\fi}}
\makeatother

\usepackage{ulem}

\def\empile#1\over#2{\mathrel{\mathop{\kern 0pt#1}\limits_{#2}}}

\def\beq{\begin{equation}}
\def\eeq{\end{equation}}
\def\bea{\begin{eqnarray}}
\def\eea{\end{eqnarray}}

\newcommand{\xbj}{$x_{{\!}_{bj}}$}

\newcommand{\Lb}{\left(}
\newcommand{\Rb}{\right)}
\def\p{{\boldsymbol p}}

\def\d3p{\frac{d^3\p}{(2\pi)^3}E_\p}

\catcode`\@=11

\newcount\@tempcntc
\def\@citex[#1]#2{\if@filesw\immediate\write\@auxout{\string\citation{#2}}\fi
  \@tempcnta\z@\@tempcntb\m@ne\def\@citea{}\@cite{%
        \@for\@citeb:=#2\do%
    {\@ifundefined{b@\@citeb}%
        {\@citeo\@tempcntb\m@ne\@citea%
                \def\@citea{,\penalty\@m\ }{\bf ?}\@warning%
                {Citation `\@citeb' on page \thepage \space undefined}}%
        {\setbox\z@\hbox{\global\@tempcntc0\csname b@\@citeb\endcsname\relax}
     \ifnum\@tempcntc=\z@ \@citeo\@tempcntb\m@ne%
       \@citea\def\@citea{,\penalty\@m}%
       \hbox{\csname b@\@citeb\endcsname}%
     \else%
      \advance\@tempcntb\@ne%
      \ifnum\@tempcntb=\@tempcntc%
      \else\advance\@tempcntb\m@ne\@citeo%
      \@tempcnta\@tempcntc\@tempcntb\@tempcntc\fi\fi}}\@citeo}{#1}}%

\def\@citeo{\ifnum\@tempcnta>\@tempcntb\else\@citea
  \def\@citea{,\penalty\@m}%
  \ifnum\@tempcnta=\@tempcntb\the\@tempcnta\else
   {\advance\@tempcnta\@ne\ifnum\@tempcnta=\@tempcntb \else
\def\@citea{--}\fi
    \advance\@tempcnta\m@ne\the\@tempcnta\@citea\the\@tempcntb}\fi\fi}

\catcode`\@=12

\begin{document}

\newcommand*{\IRFU}{IRFU, CEA, Universit\'e Paris-Saclay, F-91191 Gif-sur-Yvette, France
}\affiliation{\IRFU}
\newcommand*{\CNN}{Department of Physics, Christopher Newport University, Newport News, VA 23606, USA
}\affiliation{\CNN}
\newcommand*{\JLAB}{Jefferson Lab, Newport News, Virginia 23606, USA
}\affiliation{\JLAB}
\newcommand*{\UNICA}{Università di Cagliari and INFN Sezione di Cagliari, I-09042 Monserrato (CA), Italy
}\affiliation{\UNICA}
\newcommand*{\DUKE}{Physics Department, Duke University, Durham, NC 27708, USA
}\affiliation{\DUKE}
\newcommand*{\TRI}{Triangle Universities Nuclear Laboratory, Durham, NC 27708, USA
}\affiliation{\TRI}
\newcommand*{\IND}{Physics Department, Indiana University, Bloomington, IN 47405, USA
}\affiliation{\IND}
\newcommand*{\MIS}{Department of Physics and Astronomy, Mississippi State University, Starkville, MS 39762, USA
}\affiliation{\MIS}
\newcommand*{\PSU}{Division of Science, Penn State University Berks, Reading, Pennsylvania 19610, USA}
\affiliation{\PSU}
\newcommand*{\MIL}{Università degli Studi di Milano and INFN Sezione di Milano, I-20133 Milano, Italy
}\affiliation{\MIL}
\newcommand*{\SYR}{Department of Physics, Syracuse University, Syracuse, NY 13244, USA}
\affiliation{\SYR}

\preprint{JLAB-PHY-26-4389}

\title{Study of SIDIS Unpolarized Cross Sections from a $^3$He Target \\
with the Solenoidal Large Intensity Device at JLab}

%
\author{Matteo Cerutti}
\email{matteo.cerutti@cea.fr}
\affiliation{\IRFU}\affiliation{\CNN}\affiliation{\JLAB}%
\author{Jian-Ping Chen}
\email{jpchen@jlab.org}
\affiliation{\JLAB}
\author{Umberto D'Alesio}
\email{umberto.dalesio@ca.infn.it}
\affiliation{\UNICA}
\author{Haiyan Gao}
\email{haiyan.gao@duke.edu}
\affiliation{\DUKE}
\affiliation{\TRI}
\author{Shuo Jia}
\email{shuo.jia@duke.edu}
\affiliation{\DUKE}
\affiliation{\TRI}
\author{\\
Vladimir Khachatryan}
\email{vlakhach@iu.edu}
\affiliation{\IND}
\affiliation{\DUKE}
\affiliation{\TRI}
\affiliation{\MIS}
\author{Alexei Prokudin}
\email{prokudin@jlab.org}
\affiliation{\PSU}
\affiliation{\JLAB}
\author{Lorenzo Rossi}
\email{lorenzo.rossi3@unimi.it}
\affiliation{\MIL}
\author{Ye Tian}
\altaffiliation[]{}
\email{tianye8001@gmail.com}
\affiliation{\SYR}
\author{Zhiwen Zhao}
\altaffiliation[]{}
\email{zwzhao@jlab.org}
\affiliation{\DUKE}
\affiliation{\TRI}
%


\date{\today}

\begin{abstract}
In this paper we present a detailed impact study of semi-inclusive deep inelastic scattering unpolarized cross sections' measurements using the proposed SoLID apparatus at Jefferson Lab.
This type of data, collected at large Bjorken \xbj, moderate values of $Q^2$ and small values of the transverse momentum of produced hadrons, $P_{hT}$, allows to study transverse momentum 
dependent (TMD) parton distribution and fragmentation functions in a still poorly explored region. We present the projected results for charged light mesons based on simulated data. For 
the azimuthal-angle integrated cross sections we adopt the TMD framework up to the next-to-next-to-next-to-leading-logarithmic (N3LL) accuracy, while a simpler TMD parton model is employed 
for the study of azimuthal angular dependencies.
\end{abstract}

\maketitle


\section{Introduction}
\label{sec:Introduction}

A considerable amount of our knowledge about the quark-gluon composition of nucleons comes from  intensive and detailed experimental and theoretical studies of the parton 
distribution functions (PDFs) \cite{Ethier:2020way} and fragmentation functions (FFs) \cite{Metz:2016swz}. Their precise knowledge is essential for a better comprehension 
of Quantum Chromodynamics (QCD), the theory of strong interactions, and of the partonic structure of the nucleon and hadronization mechanism. Furthermore, a deep understanding 
of QCD is essential to search for new physics beyond the Standard Model.

For quarks collinear to the parent hadron, the cross section for the deep inelastic scattering (DIS)~\cite{Accardi:2008ne} process, can be written in terms of 
collinear PDFs. At the lowest order, these PDFs have a clear physical interpretation as probability densities for finding an unpolarized parton in a fast-moving unpolarized 
nucleon, carrying a light-cone 
momentum fraction $x$. The integrated FFs, in turn, describe the hadronization process, namely how quarks and gluons transform into color-neutral hadrons. These are, by definition, 
one-dimensional (1D) maps of the internal dynamics of nucleons in momentum space. Moreover, the PDFs depend on a resolution (factorization) scale, associated with the hard scale 
$Q^{2}$ (four-momentum transfer squared) of the respective lepton-nucleon scattering process. Thanks to more than fifty years of dedicated efforts in their theoretical and experimental 
investigations, those PDFs are known at a very high accuracy~\cite{Ball:2022qtp,Bailey:2020ooq,Hou:2019efy}. 

However, over more than two decades, the frontier of PDF exploration has been extended to the three-dimensional (3D) momentum space. Such a 3D framework has been worked out in detail 
for semi-inclusive deep inelastic scattering (SIDIS) of a lepton scattered off a nucleon, in which the leading hadron is also detected and its transverse momentum, $P_{hT}$, is 
measured~\cite{Boglione:2019nwk,Bacchetta:2006tn}. In order to describe such a process, in the small $P_{hT}$ region, one has to necessarily include the parton (quark) intrinsic transverse 
momentum $\boldsymbol{k}_{\perp}$ in the scattering picture. More precisely, the regime in which the process can be described using transverse momentum dependent (TMD) factorization theorem, 
see e.g. Ref.~\cite{Bacchetta:2006tn,Boussarie:2023izj}, corresponds to  $q_T^2 \simeq P_{hT}^2 / z^2 \ll Q^2$, where $z$ is the light-cone fraction of the fragmenting quark momentum carried 
by the produced hadron, $q_T$ is the transverse momentum of the exchanged virtual boson in a frame where the two hadrons are collinear. In TMD factorization the SIDIS cross section is given 
in terms of convolution of TMD PDFs and TMD FFs~\cite{Boer:1997nt,Goeke:2005hb,Bacchetta:2006tn,Aybat:2011zv,Angeles-Martinez:2015sea,Barone:2015ksa,Anselmino:2011,Bastami:2018xqd,Metz:2016swz}
 
Proper combinations of the lepton beam and nucleon target polarizations in SIDIS experiments give rise to azimuthal modulations that include pivotal information on various TMDs~\cite{Bacchetta:2006tn}. 
These eventually allow us to quantify the quark transverse motion inside the nucleon, observe spin-spin and spin-orbit correlations, as well as obtain quantitative insight on the 
quark orbital angular momentum (OAM)~\cite{Bacchetta:2009yq}. It is important to stress that in order to access TMDs and have a complete 3D picture of the nucleon and of the fragmentation mechanism,  
measurements from different processes are essential: that is SIDIS from HERMES, COMPASS and JLab experiments~\cite{HERMES:2015,Goertz:2002vv,JeffersonLabHallA:2011ayy,JeffersonLabHallA:2011vwy}, 
together with other processes such as Drell-Yan~\cite{Boer:2006eq,Arnold:2008kf,Collins:1984kg} and $e^{+}e^{-}$ annihilation~\cite{Metz:2016swz,Bacchetta:2015ora,Collins:1981uk}.

The SoLID Collaboration has meticulously designed and developed the concept of a large acceptance spectrometer (detector system) capable of handling high 
luminosities~\cite{JeffersonLabSoLID:2022iod,preCDR:2019,Chen:2014psa}. The proposed rich and vibrant SoLID science program will take advantage of the full potential of the JLab 
12~GeV energy upgrade~\cite{Dudek:2012vr,Arrington:2022}. There will be a new kinematic window for accomplishing precision studies of the transverse spin and TMD structure in the valence quark 
region for both neutron and proton. The experimental program on TMDs is considered one of the main pillars of the 12~GeV physics era at JLab. 

The SIDIS part of the SoLID program includes three experiments on azimuthal single- and double-target spin asymmetry measurements, using transversely (E12-10-006 \cite{E12-10-006,Gao:2010av}) 
and longitudinally (E12-11-007 \cite{E12-11-007}) polarized ${}^{3}$He (as effective polarized neutron) targets, as well as a transversely (E12-11-108 \cite{E12-11-108}) polarized NH$_{3}$ 
(proton) target. To extract TMDs with high precision from these azimuthal asymmetry measurements, the detection system of SoLID will have a capability for handling large luminosities, a full 
($2\pi$) azimuthal angular coverage, good kinematic coverage in terms of \xbj (the Bjorken $x$), $Q^{2}$ (the virtuality of the exchanged photon), $P_{hT}$ (the transverse momentum of the detected hadron), $z_{h}$ (the scaling energy fraction of the produced hadron), and good PIDs (particle identification) for electrons and charged pions. Such a setup will allow for measurements in 
multi-dimensional (4D) kinematic bins with high statistics and well-controlled systematics. 

Although azimuthal asymmetries provide information on spin-independent and spin-dependent TMD  effects, mainly differential hadron multiplicity distributions (ratios between SIDIS and DIS cross 
sections) have been widely used in state-of-the-art extractions of unpolarized TMDs~\cite{Scimemi:2019cmh,Bacchetta:2022awv,Bacchetta:2024qre,Moos:2025sal}. In contrast, there are limited studies 
on SIDIS unpolarized (and polarized) cross sections by themselves. Currently, the only available measurements of multiplicities are those from HERMES and COMPASS Collaborations~\cite{HERMES:2012uyd,COMPASS:2017mvk,CLAS:2008nzy}. 
While these measurements enabled such pioneering investigations, to study both the shape and the magnitude of the SIDIS cross sections and ascertain the validity of the factorization 
theorems, the information from absolute cross-section measurements is mandatory. The main reason is that while multiplicities, being {\it ratios} of differential cross sections, are 
certainly better under control from the experimental point of view (cancellation of systematics, etc.), they present some drawbacks from the theoretical point view, which can be summarized 
as follows: $(i)$ the denominator of a multiplicity is an inclusive quantity, namely the inclusive DIS cross section, summed over all final-state hadrons, integrated over their kinematical 
variables and given in terms of ordinary structure functions, nevertheless, its numerator is less inclusive and strongly dependent on the kinematical cuts; $(ii)$ these two factors 
entering the multiplicity are computed in different factorization schemes: its denominator in the collinear approach, while the numerator within a TMD approach; $(iii)$ the assumptions 
and the approximations adopted in these two schemes (like the role played by higher-twist contributions in the numerator) are somehow different and independent. In other words, by 
integrating the numerator entering the multiplicity over all final-state hadronic variables (and summing over all hadrons), one does not necessarily recover its denominator.

All of this eventually implies that any discrepancy between experimental data and theoretical estimates can be ascribed to different issues, which are not straightforward to 
disentangle, including also what has been used for the denominator. This might be one of the reasons of the normalization problems encountered in past analyses~\cite{Bacchetta:2022awv,Bacchetta:2024qre,OsvaldoGonzalez-Hernandez:2019iqj}. In contrast, 
a direct comparison of absolute cross-section data with estimates obtained within a TMD factorization scheme allows for a direct and unbiased access to the non-perturbative 
parts of the TMDs (as the main goal), as well as for a test of certain assumptions adopted in this scheme and/or of the potential role of higher-twist contributions.

Since the SoLID projections we are going to show are obtained for a $^3$He target, a comment on potential nuclear effects is mandatory. In this respect, currently their proper and 
consistent implementation within a TMD factorization approach is not yet fully developed (see Ref.~\cite{Alrashed:2023xsv} for pioneering studies). This is the main reason why, 
in the following, we adopt a simple model (in Sec.~\ref{sec:phi-dependent}) based on isospin symmetry, without including any dilution factor. On the other hand, by using 
measurements of absolute cross sections off $^3$He, Deuterium and Hydrogen targets, one will be able to extract important information on this factor. This could be eventually 
used for an estimate of nuclear effects and their role {\it posteriorly}, or complementarily, to tune current theoretical estimates and learn, at least indirectly, on nuclear TMDs. 
Notice that, once again, the role played by the measurements of absolute cross sections is fundamental in this respect. The use of multiplicities, as stressed above, in such a 
case would prevent any direct and unambiguous access to nuclear corrections, since many effects could mix up. Thereby, cross-section measurements will allow us to study many other 
effects that are usually neglected when working with multiplicities, such as nuclear corrections (EMC effect, nuclear binding, Fermi motion, off-shell effects), kinematic corrections, 
higher-twist effects, and so on (see e.g. Ref.~\cite{Cerutti:2025yji} and references therein). Notably, the unpolarized SIDIS cross section has already started to attract considerable 
interest almost a decade ago~\cite{Martin:2016vpv}.

The first measurements of the unpolarized cross section is about measuring a large set of SIDIS cross sections in Hall~C at JLab by the experiment E00-108 reported in \cite{Asaturyan:2011mq,Navasardyan:2006gv}. The measurements have been conducted for semi-inclusive charged-pion electro-production off proton and deuterium targets using a 5.5~GeV 
electron beam. Afterwards, a new measurement reported in \cite{JeffersonLabHallA:2016ctn,Yan:2017} has been carried out by the SIDIS experiment E06-010 in Hall~A at JLab, using a 
5.9~GeV polarized electron beam and a transversely polarized ${}^{3}$He target.  The upgraded beam energy experiment E12-09-017 (in Hall C), utilizing an 11~GeV longitudinally 
polarized electron beam scattered off proton and deuterium targets aims to determine the transverse momentum dependence of the SIDIS cross section~\cite{HallC_SIDIS_PT} with 
results forthcoming. 

The work in \cite{JeffersonLabHallA:2016ctn,Yan:2017} presents the current world data on the SIDIS unpolarized cross section, in multi-dimensional bin sets, compared with three 
theoretical models from \cite{Barone:2015ksa,Anselmino:2013lza,Bacchetta:2011gx}, whereas the target ${}^{3}$He nucleus is approximated as one neutron and two protons in a plane-wave 
scenario. Azimuthal modulations in that analysis were observed to be consistent with zero within the experimental uncertainties. Using a particular functional form, within a 
simplified leading-order (LO) TMD approach, as in the global analysis of Ref.~\cite{Barone:2015ksa}, the fitted results showed that the width of the quark intrinsic transverse momentum, 
$\langle k_{\perp}^{2} \rangle$, turned out to be much smaller than that determined from the global analyses of other types of data 
\cite{Barone:2015ksa,Anselmino:2013lza,Bacchetta:2011gx,Anselmino:2005nn}. Despite the general agreement between this simple model at the leading twist used in the analysis and the 
measured data, fairly large differences exist in some kinematic ranges, which might be related to higher-twist terms. Those terms might also be responsible for the very different 
$\langle k_{\perp}^{2} \rangle$ values determined between that study and those from Refs.~\cite{Barone:2015ksa,Anselmino:2013lza,Bacchetta:2011gx,Anselmino:2005nn}. In this respect, 
high-precision cross-section data in the moderate $Q^{2}$ range with a full azimuthal angular coverage (such as SoLID) can provide opportunities for exploring the higher-twist effects 
on azimuthal angular modulations, thereby significantly advancing our understanding of TMD physics. It is important to emphasize that theoretical proofs of factorization theorems 
exist for leading-twist observables, such as unpolarized SIDIS cross sections. Twist-3 observables are currently less studied theoretically, even though certain progress in 
next-to-leading power factorization has been achieved \cite{Bacchetta:2019qkv,Vladimirov:2021hdn,Ebert:2021jhy,Rodini:2023plb,Gamberg:2022lju}.

In this paper, we present a detailed analysis on the SIDIS unpolarized cross section measurements with the SoLID apparatus. In order to obtain physics-impact results with SoLID projections, 
we utilize a well-consolidated \texttt{MAPTMD24} framework~\cite{Bacchetta:2024qre} with TMD evolution up to next-to-next-to-next-to-leading-logarithmic (N$^3$LL) accuracy. On the 
other hand, in the study of the azimuthal modulations we will employ a simple, phenomenological LO TMD (or, generalized parton) model of Ref.~\cite{Barone:2015ksa}. Within this 
approach, TMDs will be parameterized starting from collinear distributions and assuming a transverse momentum dependence modeled as Gaussians. Although this approach is certainly 
too simple for the description of the actual experimental data, we believe that it can be very useful in exploring several issues. Studies on a variety of topics could originate when 
new data on the unpolarized cross section became available. Those may include analyses of nuclear effects in the large-$x$ region, and higher-twist corrections (still unexplored theoretically). 
For the present unpolarized cross-section studies we project the simulated SoLID pseudo-data (including the statistical and systematic uncertainties) in various combinations of 
multi-dimensional bins of up to five variables -- \xbj, $P_{hT}, z_{h}, Q^{2}, \phi_{h}$. 

The paper is organized as follows. In Sec.~\ref{sec:theory} we describe the theoretical formalism, which underlies the generation of pseudo-data obtained from a modification of the 
original SIDIS event generator from~\cite{Liu:2019}. Here we will pay great attention to the azimuthal-integrated and azimuthal-dependent cross-section scenarios. In Sec.~\ref{sec:nucl_corr} 
we discuss some results obtained from the implementation of nuclear effects (from Fermi motion) in the event generator. In Sec.~\ref{sec:res} and three appendices of this paper, we 
present the \texttt{MAPTMD24+SoLID} impact results both for the azimuthal-integrated and the azimuthal-dependent cross sections. Finally, we discuss our findings in Sec.~\ref{sec:sum}. 

More details can be found in the approved proposal of Ref.~\cite{RG_E12-10-006_12-11-007},
the proposed experiment of which will run parasitically to the SoLID SIDIS experiments E12-10-006 \cite{E12-10-006} and E12-11-007 
\cite{E12-11-007}. 

\section{SIDIS unpolarized cross section: theoretical formalism}
\label{sec:theory}


Here we present two complementary formalisms: one for the azimuthal-integrated cross sections and another one for their azimuthal angular dependencies.  
Due to the existing theoretical developments, we have to consider them separately and with different limitations/assumptions.

\subsection{Basics of the general formalism}
\label{sec:basics}

\begin{figure}[h!]
\vskip 0.0truecm
\includegraphics[width=1.0\linewidth]{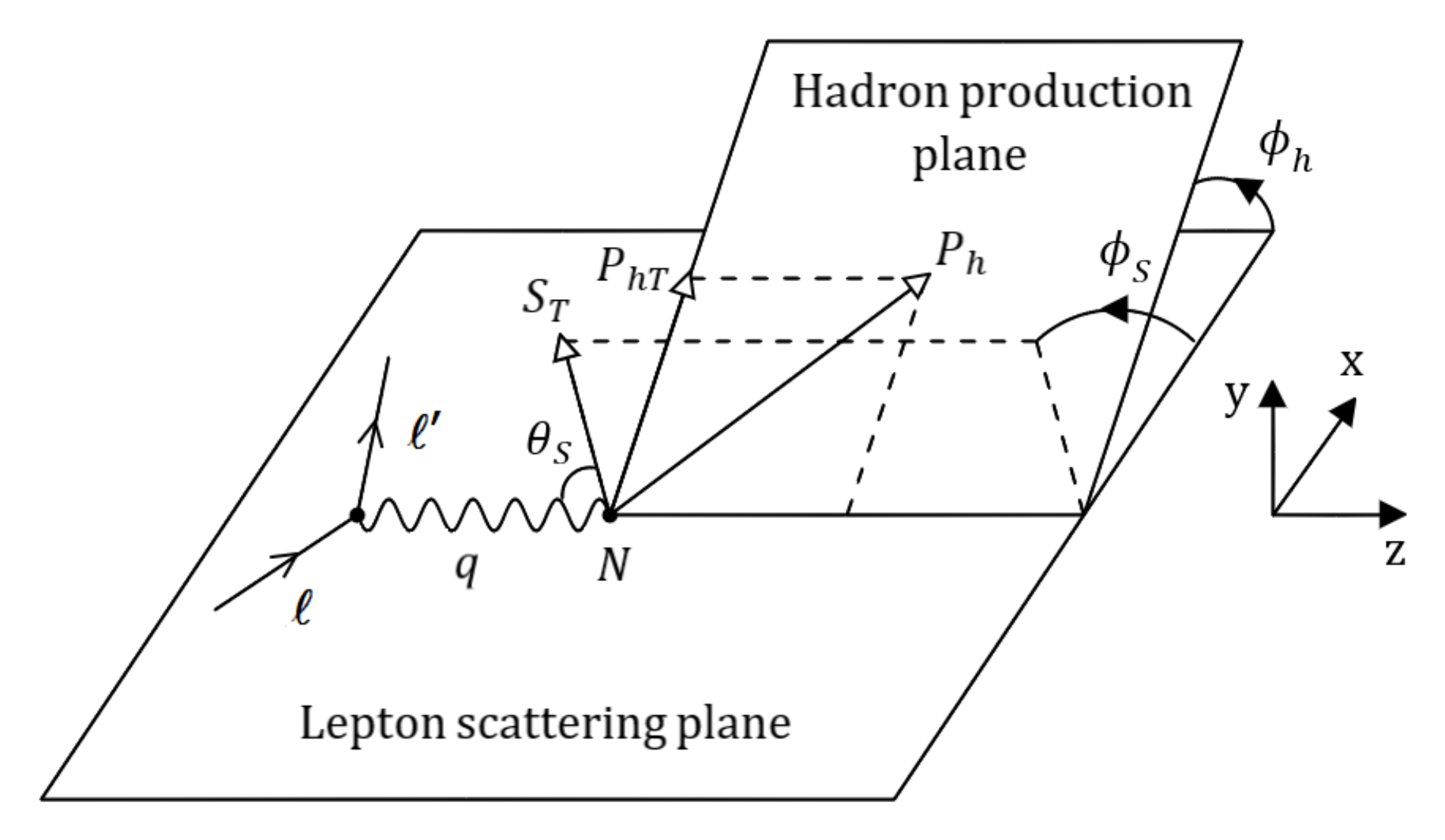}
\vskip 0.0truecm
\caption{Kinematics of the SIDIS process in Eq.~(\ref{eq:eqn_SIDIS}), represented in the one-photon exchange approximation.
The sketch follows the Trento conventions~\cite{Bacchetta:2004jz}. The $x$-$z$ plane is defined by the initial $\ell$ and 
scattered $\ell'$ lepton momenta. The azimuthal angle $\phi_{h}$ is given in the target rest frame. 
The figure is taken from Ref.~\cite{Byer:2022bqf}.}
\label{fig:fig_SIDIS}
\end{figure}

We consider the unpolarized SIDIS process in the $\gamma^{\ast}$-$N$ center-of-mass frame, where the virtual photon moves along the positive $z$ direction. 
The SIDIS process is represented as \cite{Bacchetta:2006tn,Barone:2015ksa,Anselmino:2011,Bastami:2018xqd,Bacchetta:2004jz} 
\beq
l(\ell) + N(P) \rightarrow l^{\prime}(\ell') + h(P_{h}) + X\,, 
\label{eq:eqn_SIDIS}
\eeq
where $l$ is the incident lepton, $N$ the target nucleon, $l^{\prime}$ the scattered lepton, and $h$ the produced (final-state) hadron. The corresponding four-momenta are shown 
in parentheses. In this scattering process, the nucleon and hadron masses are given by $M_{N}$ and $M_{h}$, respectively. The cross section of the process is expressed in terms 
of the following invariants:
\beq
\mbox{\xbj} = \frac{Q^{2}}{2P \!\cdot\! q} ,~~ y = \frac{P \!\cdot\! q}{P \!\cdot\! \ell} ,~~ z_{h} = \frac{P \!\cdot\! P_{h}}{P \!\cdot\! q} ,
~~ \gamma = \frac{2M_{N}\mbox{\xbj}}{Q} ,
\label{eq:eqn_invs}
\eeq
with $q\equiv \ell-\ell'$ and the hard scale $Q^{2}=-q^2$ describing the virtuality of the exchanged photon. The transverse momentum of the produced
hadron is designated as $P_{hT}=|\bm{P}_{hT}|$, as shown in Fig.~\ref{fig:fig_SIDIS}, where $\phi_{h}$ is the angle between the hadron production plane and the lepton scattering plane, 
while $\phi_{S}$ is the angle between the polarization vector of the target’s spin and the lepton scattering plane. Notice that when not specified otherwise, by $a_T$ or $a_\perp$, we 
always refer to the modulus of the corresponding vector.

In the one-photon exchange approximation, the SIDIS unpolarized differential cross section is given by \cite{Bacchetta:2006tn,Barone:2015ksa,JeffersonLabHallA:2016ctn}
\bea
\frac{d\sigma}{dx_{{\!}_{bj}} dz_{h} dQ dP_{hT}^{2}
d\phi_{h}} & = & 
\frac{4\pi \alpha^{2}}{x_{{\!}_{bj}} {Q^{3}}} \Lb 1 + 
\frac{\gamma^{2}}{2x_{{\!}_{bj}}} \Rb \bigg[ c_{1}\,F_{UU} +
\nonumber \\
& & \!\!\!\!\!\!\!\!\!\!\!\!\!\!\!\!\!\!\!\!\!\!\!\!\!\!\!\!\!\!\!\!\!\!\!\!\!\!\!\!\!\!\!\!\!
+ c_{2} \cos{\!(\phi_{h})}\,F_{UU}^{\cos{\!(\phi_{h})}}
+ c_{3} \cos{\!(2\phi_{h})}\,F_{UU}^{\cos{\!(2\phi_{h})}} \bigg] \,.
\label{eq:eqn_FUU}
\eea
In Eq.~(\ref{eq:eqn_FUU}), $\alpha$ is the electromagnetic fine-structure constant; the coefficients $c_{1}$, $c_{2}$, 
and $c_{3}$ (see Eqs.~(2.9)-(2.13) of Ref.~\cite{Bacchetta:2006tn} for more details) are given by
\bea
& & c_{1} = \frac{y^{2}}{2(1 - \varepsilon)} , \,\,\,\,\,c_{2} = \frac{y^{2}}{2(1 - \varepsilon)}\,\sqrt{2\varepsilon(1 + \varepsilon)} ,
\nonumber \\
& & \quad\quad\quad\quad\quad\quad
c_{3} = \frac{y^{2}}{2(1 - \varepsilon)}\varepsilon ,
\label{eq:eqn_factor}
\eea
where $\varepsilon$ is the ratio of longitudinal and transverse photon fluxes:
\beq
\varepsilon = \frac{1 - y - \Lb \gamma^{2}y^{2}/4 \Rb}{1 - y + \Lb y^{2}/2 \Rb + \Lb \gamma^{2}y^{2}/4 \Rb} .
\label{eq:eqn_epsil}
\eeq
The structure functions $F_{UU}$, $F_{UU}^{\cos{\!(\phi_{h})}}$, and $F_{UU}^{\cos{\!(2\phi_{h})}}$, depend on the kinematic variables \xbj, $z_{h}$, $Q^{2}$, 
and $P_{hT}^{2}$. $F_{UU}$ survives upon integration over $\phi_{h}$, while $F_{UU}^{\cos{\!(\phi_{h})}}$ and $F_{UU}^{\cos{\!(2\phi_{h})}}$ are structure 
functions related to the azimuthal $\cos{\!(\phi_{h})}$ and $\cos{\!(2\phi_{h})}$ modulations, respectively. 

As mentioned previously, we will discuss $F_{UU}$ by adopting the TMD factorization theorem and its phenomenological application at the current highest accuracy
(Sec.~\ref{sec:phi-integrated}), whereas $F_{UU}^{\cos{\!(\phi_{h})}}$ and $F_{UU}^{\cos{\!(2\phi_{h})}}$ will be studied at the parton model accuracy
(Sec.~\ref{sec:phi-dependent}).

\subsection{Azimuthal-integrated cross sections}
\label{sec:phi-integrated}

Here we discuss the structure function $F_{UU}$, adopting the TMD factorization theorem within the Collins-Soper-Sterman (CSS) framework~\cite{COLLINS1982446,Collins2013}. 
We will give only the main formulas relevant for the analysis we are interested in. All the details can be found in the literature (see, for instance, Ref.~\cite{Bacchetta:2024qre}, 
where the advanced \texttt{MAPTMD24} framework by the MAP Collaboration is presented).

\begin{figure}[h!]
\includegraphics[width=1.0\linewidth]{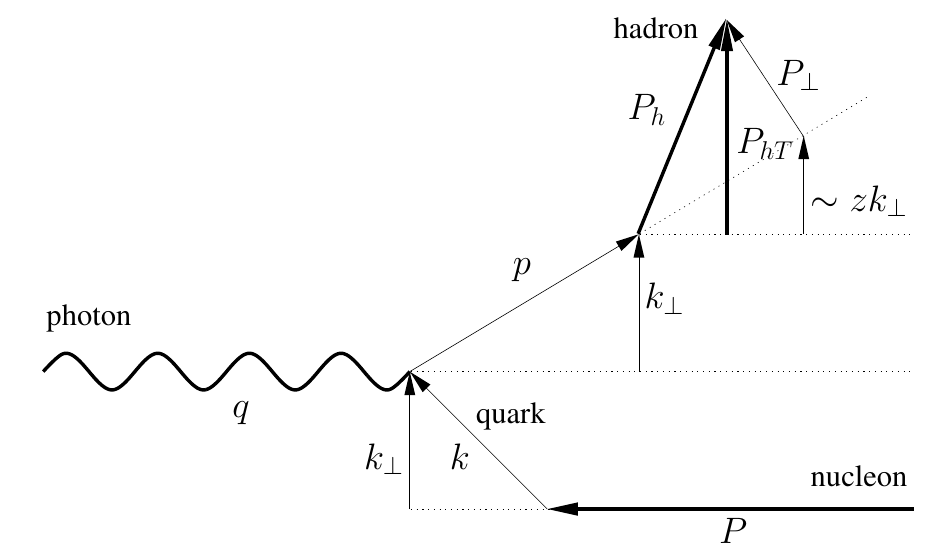}
\vskip 0.0truecm
\caption{Relevant SIDIS momenta in the $\gamma^{\ast}$-$N$ center-of-mass frame:
$q$ is the virtual photon momentum, $k$ is the momentum of the initial quark with transverse momentum $k_\perp$ inside a nucleon with momentum $P$.  
The figure is taken from \cite{Bacchetta:2024qre}, where the photon line stands for a virtual photon.}
\label{f:trans_momenta_SIDIS}
\end{figure}

A word of caution is needed on the definition of transverse momenta. Indeed, we consider the transverse component 
of the produced hadron momentum 
with respect to $P$, and $q$ ($P_{hT}$) or, equivalently, the transverse component ($q_T=|\bm{q}_T|$) of the virtual photon momentum with respect to $P$ and $P_h$ (in a frame, where 
these momenta are collinear). These are related as follows:  
\beq
q_{T}^{2} \sim P_{hT}^{2}/z_{h}^{2} .
\label{e:qT_PhT}
\eeq
The unpolarized cross section at small transverse momenta, integrated over the azimuthal angle $\phi_h$  and neglecting the target mass, can be expressed as
\begin{eqnarray*}
\frac{d\sigma^{\rm SIDIS}}{dx_{bj} dz_h dq_T dQ} & = & \frac{8 \pi^2 \alpha^2 z_h^2 q_T}{x_{bj} Q^3}
\nonumber \\
& & \!\!\!\!\!\!\!\!\!\!\!\!\!\!\!\!\!\!\!\!
\times \left[ 1+ \left( 1- \frac{Q^2}{x_{bj} s} \right)^2 \right] \, F_{UU} (x_{bj}, z_h, q_T, Q^2) , 
\end{eqnarray*}
\begin{widetext}
\begin{eqnarray}
F_{UU}(x_{bj}, z_h, q_T, Q^2) & = &  x_{bj}\,{\cal H}^{\rm SIDIS} (Q, \mu) \sum_a e_a^2 \nonumber\\
&& \times  \int\! d^2\bm{k}_\perp\! \int \!\frac{d^2 \bm{P}_\perp}{z^2}\, f_1^a (x, \bm{k}_\perp^2; \mu, \zeta_A) \,\, D_1^{a\to h} (z, \bm{P}_\perp^2; \mu, \zeta_B) \,\, 
\delta^{(2)} (\bm{k}_\perp + \bm{P}_\perp/z + \bm{q}_T) \,,
\label{eq:SIDIS_xsec}
\end{eqnarray}
\end{widetext}
\noindent where the sum in $F_{UU}$ runs over all active quark flavors. The hard factor ${\cal H}^{\text{SIDIS}}$ is perturbatively computable and depends on $Q$ and a 
renormalization scale $\mu$. The last line contains the convolution of the unpolarized TMD PDF, $f_1^a$, as a function of the rapidity scale $\zeta_{A}$ and of the transverse 
momentum $k_\perp=|\bm{k}_\perp|$ of the struck quark with respect to the nucleon momentum, as well as the TMD FF, $D_1^{a \to h}$, as a function of the rapidity scale $\zeta_{B}$ 
and of the transverse momentum $P_\perp=|\bm{P}_\perp|$ of the produced hadron $h$ with respect to the fragmenting quark axis. 
To have a better visualization, in Fig.~\ref{f:trans_momenta_SIDIS} we show the involved transverse momenta in the $\gamma^{\ast}$-$N$ center-of-mass frame.
The struck parton, carrying the four-momentum $p=k+q$, fragments into a hadron with four-momentum $P_h$ that has a transverse momentum $P_\perp$ with respect to the fragmenting quark axis. 
The total (measured) transverse momentum of the final-state hadron is $P_{hT}$, and 
in the large-$Q^2$ limit $P_{hT} \approx z k_\perp + P_\perp$.

In the following, we adopt the large-$Q$ limit, where $x=$ \xbj and $z=z_h$. As usual, one works in the conjugate position space ($\bm{b}_T$ space) by defining the Fourier transforms 
of the TMD PDF
\begin{equation}
\label{eq:FTdef}
  \begin{split}
\hat f^a_1 \big( x, \bm{b}_T^2; \mu, \zeta \big) &= \int d^2 \bm{k}_\perp \, e^{i
      \bm{b}_T \cdot \bm{k}_\perp  } \, f_1^a \big( x, \bm{k}_\perp^2; \mu,
    \zeta \big) \,
    \\
      & \!\!\!\!\!\!\!\!\!\!\!\!\!\!\!\!\!\!\!\!
      = 2 \pi \int_0^{\infty} d k_\perp \,k_\perp  J_0(b_T k_\perp) \, f_1^a \big( x, \bm{k}_\perp^2; \mu, \zeta \big) ,
 \end{split}
\end{equation}

\hskip -0.35truecm
and the TMD FF
\begin{align}
 \begin{split}
\label{eq:FTdefFF}
\hat{D}_1^{a \to h}\big( z, \bm{b}_T^2; \mu, \zeta \big) = \,
    \\
      & \!\!\!\!\!\!\!\!\!\!\!\!\!\!\!\!\!\!\!\!\!\!\!\!\!\!\!\!\!\!\!\!\!\!\!\!\!\!\!\!\!\!\!\!\!\!\!\!\!
      = \int \frac{d^2 \bm{P}_\perp}{z^2} \, e^{-i
      \bm{b}_T  \cdot  \bm{P}_\perp/z } \, D_1^a \big( z, \bm{P}_\perp^2; \mu, \zeta \big) \,
    \\
      & \!\!\!\!\!\!\!\!\!\!\!\!\!\!\!\!\!\!\!\!\!\!\!\!\!\!\!\!\!\!\!\!\!\!\!\!\!\!\!\!\!\!\!\!\!\!\!\!\!
      = 2 \pi \int_0^{\infty} \frac{d P_\perp}{z^2} \,P_\perp  J_0(b_T P_\perp/z) \, D_1^a \big( z, \bm{P}_\perp^2; \mu,
    \zeta \big) ,
 \end{split}
\end{align}
where $J_{0}$ is the Bessel function of the first kind. The convolution in the second line of Eq.~(\ref{eq:SIDIS_xsec}) can be rewritten as
\bea
\label{e:FUUT_def}
& & \!\!\!\!\!\!\!\!\!
\frac{1}{2\pi}\, \int_0^{+\infty} db_T\,b_T\,J_0\big( b_T q_T \big) \times
\nonumber \\
& & \quad\quad\quad
\times \hat{f}_1^a(x,\bm{b}_T^2;\mu,\zeta_A)\, \hat{D}_1^{a \to h}(z,\bm{b}_T^2;\mu,\zeta_B) .
\eea

Focusing on the evolution of TMD PDFs (an analogous description applies to TMD FFs as well), i.e., the dependence on the renormalization scale $\mu$ and the 
rapidity scale $\zeta$, the complete set of equations is given by
\begin{eqnarray}
\frac{\partial \ln\hat{f}_{1}(x,\bm{b}_T^2,\mu,\zeta)}{\partial\ln\mu} & = &\gamma_F\!\left(\alpha_s(\mu),\frac{\zeta}{\mu^2}\right),\nonumber\\
\frac{\partial \hat{f}_{1}(x,\bm{b}_T^2,\mu,\zeta)}{\partial\ln\sqrt{\zeta}} & = & K(b_T,\mu), \\
\frac{\partial K(b_T,\mu)}{\partial\ln\mu} & = &-\gamma_{K}(\alpha_{s}(\mu))\,,\nonumber
\label{e:evoleq}
\end{eqnarray}
where $\gamma_F$/$K$ are the anomalous dimensions of the renormalization-group and of the Collins-Soper evolution equations, respectively, while $\gamma_{K}$ is the 
so-called cusp anomalous dimension. 

These differential equations allow us to determine TMD PDFs at any final pair of scales ($\mu_f,\zeta_f$), starting from an initial pair ($\mu_i,\zeta_i$). 
In addition, in the region of small $b_T$, the TMD PDF, $\hat{f}_{1}$, is matched onto its corresponding collinear PDF, $f_1$, through a convolution with suitable 
perturbative matching coefficients $C$. The resulting expression at the final scales ($\mu_f, \zeta_f$) is
\begin{eqnarray}
& & \!\!\!\!\!\!\!\!\!\!\!\!\!\!\!\!\!
\hat{f}_{1}(x,\,\bm{b}_T^2;\,\mu_f,\zeta_f) = \big[C\otimes f_1\big] (x,\,\bm{b}_T^2;\,\mu_i,\zeta_i) \times
\nonumber \\
& & \!\!\!\!\!\!\!\!\!\!\!\!\! \times 
\exp\bigg\{ K(\mu_{i})\ln\!\biggl(\frac{\sqrt{\zeta_{f}}}{\sqrt{\zeta_{i}}}\biggr) + 
\nonumber \\
& & \!\!\!\!\!\!\!\!\!\!\!\!\!
+ \int_{\mu_i}^{\mu_f} \frac{d\mu}{\mu}\, 
\bigg[\gamma_{F}(\alpha_{s}(\mu),1)-\gamma_{K}(\alpha_{s}(\mu))\ln\!\biggl(\frac{\sqrt{\zeta_{f}}}{\mu}\biggr)\bigg]\bigg\} .
\label{e:evolved_TMDs}
\end{eqnarray}
More details can be found in Ref.~\cite{Bacchetta:2024qre} and references therein.

The initial scale $\mu_i$ is commonly chosen for simplifying the calculations of perturbative ingredients. 
More precisely, one adopts $\mu_i \sim 1/b_T$; this choice, while well under control at small $b_T$, manifests a problem when $b_T$ becomes too large: the hitting 
of the Landau pole when computing the QCD coupling constant at too low $\mu_i$. A suitable and well consolidated prescription to cure this problem within the CSS 
framework is the so-called $b_*$-prescription, by redefining $\mu_i$ as $\mu_{b*} \sim 1/b_*(b_T)$. Through this procedure, the TMD PDF in Eq.~(\ref{e:evolved_TMDs}) 
can be written as
\begin{widetext}
\bea
\label{e:TMD_pertvsNP}
\hat{f}_{1}(x,\,\bm{b}_T^2;\,\mu_f,\zeta_f) = \hat{f}_{1}(x,\,\bm{b}_T^2;\,\mu_f,\zeta_f)\bigg |_{\text{pert}} \hat{f}_{\rm NP}(x,\,\bm{b}_T^2;\,\zeta, Q_0) , 
\eea
\end{widetext}
namely, in terms of the product between a fully perturbatively calculable expression and a nonperturbative function that must be fitted to experimental data
($Q_0$ is the scale at which this latter contribution is parametrized).

In this analysis we consider the parameterization of $\hat{f}_{\rm NP}$ taken from the recent N$^3$LL extraction of \cite{Bacchetta:2024qre}, where a detailed study 
of the flavor dependence was carried out at the currently available highest theoretical accuracy. For clarity, we report the \texttt{MAPTMD24} model 
selected for the TMD PDF, $\hat{f}_{1}$:
\begin{widetext}
\beq
f_{\rm NP}(x, \bm{b}_T^2; \zeta, Q_0) =
\frac{
g_1(x)\, e^{ - g_1(x) \frac{\bm{b}_T^2}{4}} +
\lambda^2\, g_{2}^2(x)\, \bigg[ 1 - g_{2}(x) \frac{\bm{b}_T^2}{4} \bigg]\, e^{ - g_{2}(x) \frac{\bm{b}_T^2}{4}} +
\lambda_2^2\, g_{3}(x)\, e^{ - g_{3}(x) \frac{\bm{b}_T^2}{4}}
}{
g_1(x) +  \lambda^2\, g_{2}^2(x) + \lambda_2^2\, g_{3}(x)
} \bigg[ \frac{\zeta}{Q_0^2}  \bigg]^{g_K(\bm{b}_T^2)/2} \, ,
\label{e:f1NP}
\eeq
\end{widetext}
corresponding to the Fourier transform of the sum of two Gaussians and a Gaussian weighted by $\bm{k}_\perp^2$. The analogous expression of the model for the TMD FF,
$\hat{D}_1$, is
\begin{widetext}
\beq
\label{e:D1NP}
D_{1\,\rm NP}(z, \bm{b}_T^2; \zeta, Q_0) =
\frac{
g_4(z)\, e^{ - g_4(z) \frac{\bm{b}_T^2}{4z^2}} +
\frac{\lambda_F}{z^2}\, g_{5}^2(z)\, \bigg[ 1 -
g_{5}(z) \frac{\bm{b}_T^2}{4z^2} \bigg]\, e^{ - g_{5}(z) \frac{\bm{b}_T^2}{4z^2}}
}{
g_4(z) +  \frac{\lambda_F}{z^2}\, g_{5}^2(z)
} \,
\bigg[ \frac{\zeta}{Q_0^2} \bigg]^{g_K(\bm{b}_T^2)/2} ,
\eeq
\end{widetext}
corresponding to the Fourier transform of the sum of a Gaussian and a Gaussian weighted by $\bm{P}_\perp^2$. The $g_i$ functions represent the widths of the distributions and include a dependence on the variables $x$ and $z$:
\beq
g_{\{1,2,3\}}(x) = N_{\{1,2,3\}} \frac{x^{\sigma_{\{1,2,3\}}}(1-x)^{\alpha^2_{\{1,2,3\}}}}{\hat{x}^{\sigma_{\{1,2,3\}}}(1-\hat{x})^{\alpha^2_{\{1,2,3\}}}} ,
\label{e:gi_func_PDF}
\eeq
\beq
g_{\{4,5\}}(z) = N_{\{4,5\}} \frac{\left(z^{\beta_{\{1,2\}}} + \delta^2_{\{1,2\}}\right)(1 - z)^{\gamma^2_{\{1,2\}}}}{
\left(\hat{z}^{\beta_{\{1,2\}}} + \delta^2_{\{1,2\}}\right)(1 - \hat{z})^{\gamma^2_{\{1,2\}}}} .
\label{e:gi_func_FF}
\eeq
with $\hat{x} = 0.1$, $\hat{z} = 0.5$, and $N_i$ ($\mbox{with}~i$ = $1$, $2$, $3$, $4$, $5$), \, $\sigma_j$, $\alpha_j$ ($\mbox{with}~j$ = $1$, $2$, $3$),
\, $\beta_i$, $\delta_i$, $\gamma_i$ ($\mbox{with}~i$ = $1$, $2$), all being as free parameters. 
Finally, $g_{K}$ is the non perturbative part of the Collins-Soper kernel: namely, the evolution on the rapidity scale $\zeta$, and is parametrized as
\begin{equation}
g_K(\bm{b}_T^2) = - g_2^2\, \frac{\bm{b}_T^2}{2} ,
\label{e:CSkernelNP}
\end{equation}
where we set $Q_0=1$~GeV.

The \texttt{MAPTMD24} global fit, employed in this SIDIS unpolarized cross-section analysis, can be considered as the current state-of-the-art phenomenological 
extraction of TMD PDFs and TMD FFs. For completeness, we should mention that another simultaneous extraction of same quantities at the current best accuracy has 
appeared very recently in Ref.~\cite{Moos:2025sal}, namely the ART25 fit.

\subsection{Azimuthal angular-dependent cross sections}
\label{sec:phi-dependent}

We now move on to the discussion of the azimuthal dependence in the unpolarized cross section in Eq.~(\ref{eq:eqn_FUU}). In such a case, we cannot utilize the full machinery 
discussed in the previous section. Nonetheless, even by adopting a simplified TMD framework, at a lower level of accuracy, we will be able to show what one can learn from 
this study, putting some bases for future improved analyses. 

Thus, we will focus on $F_{UU}^{\cos{\!(\phi_{h})}}$ and $F_{UU}^{\cos{\!(2\phi_{h})}}$; besides, for consistency and internal coherence, we will also discuss $F_{UU}$ 
in the same framework. Hereinafter, we mainly follow Ref.~\cite{Barone:2015ksa}, showing some of its formulas, which we use in this part of our analysis for calculating 
the central values of the unpolarized cross-section pseudo-data. Consistently, we will directly work in $\bm{k}_\perp$ space.

The LO factorization scheme of Eq.~(\ref{eq:eqn_FUU}), supplemented by Eq.~(\ref{eq:eqn_factor}), is equivalent to Eq.~(3) of Ref.~\cite{{Barone:2015ksa}}, with
\beq
F_{UU} = \sum\limits_{q} e_{q}^{2}\,x \int d^{2}\bm{k}_\perp\,f_{q}(x, \bm{k}^2_{\perp}) D_{q}(z, \bm{P}^2_{\perp}) \,,
\label{eq:eqn_FUU_TMD}
\eeq
where the sum is assumed to be over quarks and antiquarks. The $Q^{2}$ dependence of both functions in the integrand is omitted for simplicity.
This enters the azimuthal-independent part of Eq.~(\ref{eq:eqn_FUU}) and it is the leading-order expression of the $F_{UU}$ in Eq.~(\ref{eq:SIDIS_xsec}), after applying 
the momentum conservation condition
\begin{equation}
\bm{P}_{hT} = z\bm{k}_{\perp} + \bm{P}_{\perp}\,. 
\label{eq:eqn_mom_cons}
\end{equation}
The second structure function in Eq.~(\ref{eq:eqn_FUU}), $F_{UU}^{\cos{\!(\phi_{h})}}$, associated to the $\cos{\!(\phi_{h})}$ modulation of the cross section, 
is a twist-3 quantity of the order of $1/Q$. It can be written as the sum of the Cahn and Boer-Mulders contributions:
\beq
F_{UU}^{\cos{\!(\phi_{h})}} = F_{UU}^{\cos{\!(\phi_{h})}}\big|_{\rm Cahn} + F_{UU}^{\cos{\!(\phi_{h})}}\big|_{\rm BM} .
\label{eq:eqn_FUU_cos}
\eeq
Herein we have
\begin{widetext}
\beq
F_{UU}^{\cos{\!(\phi_{h})}}\big|_{\rm Cahn} =  -2 \sum\limits_{q} e_{q}^{2}\,x \int d^{2}\bm{k}_\perp\,\frac{(\bm{k}_\perp\!\cdot\!\hat{\bm{h}})}{Q}\,f_{q}(x, \bm{k}^2_{\perp}) D_{q}(z, \bm{P}^2_{\perp}) ,
\label{eq:eqn_FUU_Cahn}
\eeq
\end{widetext}
\begin{widetext}
\beq
F_{UU}^{\cos{\!(\phi_{h})}}\big|_{\rm BM} = \sum\limits_{q} e_{q}^{2}\,x \int d^{2}\bm{k}_\perp\,\frac{k_{\perp}}{Q}\frac{P_{hT} - z\,(\bm{k}_\perp\!\cdot\!\hat{\bm{h}})}{k_{\perp}} 
\,\Delta f_{q^{\uparrow}/p}(x, \bm{k}^2_{\perp})\,\Delta D_{h/q^{\uparrow}}(z, \bm{P}^2_{\perp}) \,,
\label{eq:eqn_FUU_BM}
\eeq
\end{widetext}
where $\hat{\bm{h}} \equiv \bm{P}_{hT}/|\bm{P}_{hT}|$ is a unit vector. The Cahn contribution in Eq.~(\ref{eq:eqn_FUU_Cahn}) involves the convolution of 
$f_{q}(x, \bm{k}^2_{\perp})$ and $D_{q}(z, \bm{P}^2_{\perp})$. The Boer-Mulders contribution in Eq.~(\ref{eq:eqn_FUU_BM}) involves the convolution of the 
Boer-Mulders TMD $\Delta f_{q^{\uparrow}/p}(x, \bm{k}^2_{\perp})$ and the Collins FF $\Delta D_{h/q^{\uparrow}}(z, \bm{P}^2_{\perp})$. These two functions 
are in turn represented by the following expressions:
\beq
\Delta f_{q^{\uparrow}/p}(x, \bm{k}^2_{\perp}) = -\frac{k_{\perp}}{M_{p}}\,h_{1}^{\perp}(x, \bm{k}^2_{\perp}) ,
\label{eq:eqn_Delta_f}
\eeq
\beq
\Delta D_{h/q^{\uparrow}}(z, \bm{P}^2_{\perp}) = \frac{2P_{\perp}}{zM_{h}}\,H_{1}^{\perp}(z, \bm{P}^2_{\perp}) ,
\label{eq:eqn_Delta_D}
\eeq
according to the Amsterdam notation \cite{Bacchetta:2006tn}. The function $h_{1}^{\perp}$ describes a quark's transverse polarization asymmetry inside an unpolarized nucleon 
\cite{Boer:1997nt}. The function $H_{1}^{\perp}$  describes the asymmetric distribution in the fragmentation of a transversely polarized quark into an unpolarized hadron
\cite{Bacchetta:2001di}. 

The third structure function in Eq.~(\ref{eq:eqn_FUU}), $F_{UU}^{\cos{\!(2\phi_{h})}}$, associated to the $\cos{\!(2\phi_{h})}$ modulation of the cross section, consists 
of a twist-4 Cahn contribution and a twist-2 Boer-Mulders term: 
\beq
F_{UU}^{\cos{\!(2\phi_{h})}} \approx F_{UU}^{\cos{\!(2\phi_{h})}}\big|_{\rm Cahn} + F_{UU}^{\cos{\!(2\phi_{h})}}\big|_{\rm BM} ,
\label{eq:eqn_FUU_cos2}
\eeq
where
\begin{widetext}
\beq
F_{UU}^{\cos{\!(2\phi_{h})}}\big|_{\rm Cahn} =  2 \sum\limits_{q} e_{q}^{2}\,x \int d^{2}\bm{k}_\perp\,\frac{2(\bm{k}_\perp\!\cdot\!\hat{\bm{h}})^{2} - \bm{k}_{\perp}^{2}}{Q^{2}} 
\,f_{q}(x, \bm{k}^2_{\perp}) D_{q}(z, \bm{P}^2_{\perp}) ,
\label{eq:eqn_FUU_Cahn2}
\eeq
\end{widetext}
\begin{widetext}
\beq
F_{UU}^{\cos{\!(2\phi_{h})}}\big|_{\rm BM} = -\sum\limits_{q} e_{q}^{2}\,x \int d^{2}\bm{k}_\perp\,\frac{P_{hT}(\bm{k}_\perp\!\cdot\!\hat{\bm{h}}) + 
z\left[\bm{k}_{\perp}^{2} - 2 (\bm{k}_\perp\!\cdot\!\hat{\bm{h}})^{2} \right]}{2k_{\perp}p_{\perp}}\,
\Delta f_{q^{\uparrow}/p}(x, \bm{k}^2_{\perp})\,\Delta D_{h/q^{\uparrow}}(z, \bm{P}^2_{\perp}) .
\label{eq:eqn_FUU_BM2}
\eeq
\end{widetext}
In Eq.~(\ref{eq:eqn_FUU_Cahn2}), light-cone parameters $x$ and $z$ do not exactly coincide with \xbj and $z_{h}$, and that formula should be considered as an approximation to the 
full twist-4 contribution to $F_{UU}^{\cos{\!(2\phi_{h})}}$.  

For the TMD parametrizations one often uses a Gaussian ansatz, as also supported by phenomenological analyses in \cite{Schweitzer:2010tt}, in which $f_{q}(x, \bm{k}_{\perp}^2)$ 
and $D_{q}(z, \bm{P_{\perp}}^2)$ are expressed as
\beq
f_{q}(x, \bm{k}^2_{\perp}) = f_{q}(x)\,\frac{ e^{-k_{\perp}^{2}/\langle k_{\perp}^{2} \rangle} }{ \pi \langle k_{\perp}^{2} \rangle } ,
\label{eq:eqn_fq}
\eeq
\beq
D_{q}(z, \bm{P}^2_{\perp}) =  D_{q}(z)\,\frac{ e^{-P_{\perp}^{2}/\langle P_{\perp}^{2} \rangle} }{ \pi \langle P_{\perp}^{2} \rangle } ,
\label{eq:eqn_Dq}
\eeq
where $f_{q}(x)$ and $D_{q}(z)$ are the collinear PDFs and FFs respectively. 

It is important to emphasize that the Gaussian widths $\langle k_\perp^2\rangle$ and $\langle P_\perp^2\rangle$ may have kinematic dependence on $x$ or $z$, as suggested 
in the parameterizations shown in Eqs.~(\ref{e:gi_func_PDF}) and (\ref{e:gi_func_FF}) and in many other TMD fits. Similarly their flavor dependence, neglected in 
Ref.~\cite{Barone:2015ksa} as in most phenomenological studies, could play a role~\cite{Bacchetta:2024qre}. Unfortunately, the scarcity of experimental data on the azimuthal 
dependence of the unpolarized cross sections, together with the more complicated theoretical framework needed to address it, prevent from reaching an analysis at the same 
level of sophistication as the one presented in Sec.~\ref{sec:phi-integrated}. 

In the following we use the LHAPDF CJ15lo set for the collinear PDF~\cite{CJ15lo}, and the DSSFFlo set for the collinear FF~\cite{FDSS}, both employed in \cite{Liu:2019} 
for $A_{UT}$ asymmetry studies of the SoLID Preliminary Conceptual Design Report in \cite{preCDR:2019}. 

The Boer-Mulders TMD function is parametrized as follows:
\bea
& & \!\!\!\!\!\!\!\!\!\!
\Delta f_{q^{\uparrow}/p}(x, \bm{k}^2_{\perp}) = 
\nonumber \\
& &
= \Delta f_{q^{\uparrow}/p}(x)\,\sqrt{2e}\,\frac{k_{\perp}}{M_{{\!}_{\rm BM}}}\,e^{-k_{\perp}^{2}/M_{{\!}_{\rm BM}}^{2}}\,
\frac{ e^{-k_{\perp}^{2}/\langle k_{\perp}^{2} \rangle} }{ \pi \langle k_{\perp}^{2} \rangle } ,
\label{eq:eqn_Delta_f2}
\eea
with
\bea
& & \!\!\!\!\!\!\!\!\!\!
\Delta f_{q^{\uparrow}/p}(x) = 
\nonumber \\
& &
= N_{q}\,\frac{\Lb \alpha_{q} + \beta_{q} \Rb^{(\alpha_{q} + \beta_{q})}}{\alpha_{q}^{\alpha_{q}} \beta_{q}^{\beta_{q}}}\,
x^{\alpha_{q}} (1 - x)^{\beta_{q}} f_{q}(x) ,
\label{eq:eqn_Delta_f22}
\eea
where $N_{q}$, $\alpha_{q}$, $\beta_{q}$, and $M_{{\!}_{\rm BM}}$ are the parameters of the Boer-Mulders quark distributions 
\cite{Barone:2009hw} (see also table~4 in \cite{Bastami:2018xqd}). By imposing $|N_q|\le 1$, the so-called positivity bound --
$|\Delta f_{q^{\uparrow}/p}(x, \bm{k}^2_{\perp})| \leq f_{q/p}(x, \bm{k}^2_{\perp})$ -- is satisfied.
The Boer-Mulders function can be rewritten also as
\bea
\Delta f_{q^{\uparrow}/p}(x, \bm{k}^2_{\perp}) & = & 
\nonumber \\
& & \!\!\!\!\!\!\!\!\!\!\!\!\!\!\!\!\!\!\!\!\!\!\!\!\!\!\!\!\!\!\!\!\!\!\!\!\!\!\!\!
 = \Delta f_{q^{\uparrow}/p}(x)\,\sqrt{2e}\,\frac{k_{\perp}}{M_{{}_{\rm BM}}}\,
\frac{ e^{-k_{\perp}^{2}/\langle k_{\perp}^{2} \rangle_{{}_{\rm BM}}} }{ \pi \langle k_{\perp}^{2} \rangle } ,
\label{eq:eqn_Delta_f3}
\eea
with 
\beq
\langle k_{\perp}^{2} \rangle_{{}_{\rm BM}} = \frac{\langle k_{\perp}^{2} \rangle\,M_{{}_{\rm BM}}^{2}}{\langle k_{\perp}^{2} \rangle +  M_{{}_{\rm BM}}^{2}} .
\label{eq:eqn_kperp_BM}
\eeq
The Collins FF is parametrized as follows:
\bea
& & \!\!\!\!\!\!\!\!\!\!
\Delta D_{h/q^{\uparrow}}(z, \bm{P}^2_{\perp}) =
\nonumber \\
& &
= \Delta D_{h/q^{\uparrow}}(z)\,\sqrt{2e}\,\frac{P_{\perp}}{M_{\rm C}}\,e^{-P_{\perp}^{2}/M_{\rm C}^{2}}\,
\frac{ e^{-P_{\perp}^{2}/\langle P_{\perp}^{2} \rangle} }{ \pi \langle P_{\perp}^{2} \rangle } ,
\label{eq:eqn_Delta_D2}
\eea
with
\beq
\Delta D_{h/q^{\uparrow}}(z) = 2 N_{q}^{\rm C}\,\frac{(\gamma + \delta)^{\gamma + \delta}}{\gamma^{\gamma}\,\delta^{\delta}}\,z^{\gamma}(1 - z)^{\delta}D_{q}(z) ,
\label{eq:eqn_Delta_D22}
\eeq
where $N_{q}^{\rm C}$, $\gamma$, $\delta$, and $M_{\rm C}$ are free parameters for the favored and disfavored Collins FFs~\cite{Anselmino:2013vqa} 
(see also table~3 in \cite{Bastami:2018xqd}). Even in this case, by imposing $|N_q^{\rm C}|\le 1$, the corresponding positivity bound for the Collins
function is fulfilled.  
 
By combining the two Gaussians in Eq.~(\ref{eq:eqn_Delta_D2}) we have:
\beq
\!\!\!\!\Delta D_{h/q^{\uparrow}}(z, \bm{P}^2_{\perp}) = \Delta D_{h/q^{\uparrow}}(z)\,\sqrt{2e}\,\frac{P_{\perp}}{M_{\rm C}}\,
\frac{ e^{-P_{\perp}^{2}/\langle P_{\perp}^{2} \rangle_{{}_{\rm C}} } }{ \pi \langle P_{\perp}^{2} \rangle } ,
\label{eq:eqn_Delta_D3}
\eeq
with 
\beq
\langle P_{\perp}^{2} \rangle_{{}_{\rm C}} = \frac{\langle P_{\perp}^{2} \rangle\,M_{\rm C}^{2}}{\langle P_{\perp}^{2} \rangle +  M_{\rm C}^{2}} .
\label{eq:eqn_pperp_C}
\eeq

These Gaussian parametrizations allow one to carry out all the integrals over the transverse momenta analytically. 
Thereby, by inserting the above distribution and fragmentation functions into the expressions of the structure functions
$F_{UU}$ in Eq.~(\ref{eq:eqn_FUU_TMD}), 
$F_{UU}^{\cos{\!(\phi_{h})}}\big|_{\rm Cahn}$ in Eq.~(\ref{eq:eqn_FUU_Cahn}), 
$F_{UU}^{\cos{\!(\phi_{h})}}\big|_{\rm BM}$ in Eq.~(\ref{eq:eqn_FUU_BM}), 
$F_{UU}^{\cos{\!(2\phi_{h})}}\big|_{\rm Cahn}$ in Eq.~(\ref{eq:eqn_FUU_Cahn2}), 
and $F_{UU}^{\cos{\!(2\phi_{h})}}\big|_{\rm BM}$ in Eq.~(\ref{eq:eqn_FUU_BM2}), 
we obtain the following set of equations (where we use $x=$ \xbj\, and $z=z_{h}$, valid within the approximations we consider):

\beq
F_{UU} = \sum\limits_{q}e_{q}^{2}\,x_{{\!}_{bj}}\,f_{q}(x_{{\!}_{bj}})\,D_{q}(z_{h})\,\frac{ e^{-P_{hT}^{2}/\langle P_{hT}^{2} \rangle} }{\pi \langle P_{hT}^{2} \rangle} ,
\label{eq:eqn_FUU_TMD_final}
\eeq
\begin{widetext}
\bea
F_{UU}^{\cos{\!(\phi_{h}})}\big|_{\rm Cahn} = -2\,\frac{P_{hT}}{Q} \sum\limits_{q} e_{q}^{2}\,x_{{\!}_{bj}}\,f_{q}(x_{{\!}_{bj}})\,D_{q}(z_{h})
\times \,\frac{ z_{h}\langle k_{\perp}^{2} \rangle}{\langle P_{hT}^{2} \rangle}\,\frac{ e^{-P_{hT}^{2}/\langle P_{hT}^{2} \rangle} }{\pi \langle P_{hT}^{2} \rangle} ,
\label{eq:eqn_FUU_Cahn_final}
\eea
\bea
& & \!\!\!\!\!\!\!\!\!\!\!\!\!\!\!\!\!
F_{UU}^{\cos{\!(\phi_{h}})}\big|_{\rm BM} = 2e\,\frac{P_{hT}}{Q} \sum\limits_{q} e_{q}^{2}\,x_{{\!}_{bj}}\,\frac{\Delta f_{q^{\uparrow}/p}(x_{{\!}_{bj}})}{M_{{}_{\rm BM}}}
\frac{\Delta D_{h/q^{\uparrow}}(z_{h})}{M_{\rm C}}\,\frac{ e^{-P_{hT}^{2}/\langle P_{hT}^{2} \rangle_{{}_{\rm BM}}} }{\pi \langle P_{hT}^{2} \rangle_{{}_{\rm BM}}^{4}}
\frac{ \langle k_{\perp}^{2} \rangle_{{}_{\rm BM}}^{2}\,\langle P_{\perp}^{2} \rangle_{{}_{\rm C}}^{2} }
{ \langle k_{\perp}^{2} \rangle \langle P_{\perp}^{2} \rangle }
\nonumber \\
& & \qquad\qquad
\times \left[ z_{h}^{2} \langle k_{\perp}^{2} \rangle_{{}_{\rm BM}} \Lb P_{hT}^{2} - \langle P_{hT}^{2} \rangle_{{}_{\rm BM}} \Rb + 
\langle P_{\perp}^{2} \rangle_{{}_{\rm C}} \langle P_{hT}^{2} \rangle_{{}_{\rm BM}} \right] , 
\label{eq:eqn_FUU_BM_final}
\eea
\bea
F_{UU}^{\cos{\!(2\phi_{h})}}\big|_{\rm Cahn} = 2\,\frac{P_{hT}^{2}}{Q^{2}} \sum\limits_{q} e_{q}^{2}\,x_{{\!}_{bj}}
f_{q}(x_{{\!}_{bj}})\,D_{q}(z_{h})\,
\frac{ z_{h}^{2}\langle k_{\perp}^{2} \rangle^{2}}{\langle P_{hT}^{2} \rangle^{2}}\,\frac{ e^{-P_{hT}^{2}/\langle P_{hT}^{2} \rangle} }{\pi \langle P_{hT}^{2} \rangle} ,
\label{eq:eqn_FUU_Cahn2_final}
\eea
\bea
F_{UU}^{\cos{\!(2\phi_{h})}}\big|_{\rm BM} = - e P_{hT}^{2} \sum\limits_{q} e_{q}^{2}\,x_{{\!}_{bj}}\,\frac{\Delta f_{q^{\uparrow}/p}(x_{{\!}_{bj}})}{M_{{}_{\rm BM}}}
\frac{\Delta D_{h/q^{\uparrow}}(z_{h})}{M_{\rm C}}\,\frac{ e^{-P_{hT}^{2}/\langle P_{hT}^{2} \rangle_{{}_{\rm BM}}} }{\pi \langle P_{hT}^{2} \rangle_{{}_{\rm BM}}^{3}}
\frac{ z_{h} \langle k_{\perp}^{2} \rangle_{{}_{\rm BM}}^{2}\,\langle P_{\perp}^{2} \rangle_{{}_{\rm C}}^{2} }{ \langle k_{\perp}^{2} \rangle \langle P_{\perp}^{2} \rangle } ,
\label{eq:eqn_FUU_BM2_final}
\eea
\end{widetext}
where
\bea
\langle P_{hT}^{2} \rangle & = & \langle P_{\perp}^{2} \rangle + z_{h}^{2}\,\langle k_{\perp}^{2} \rangle ,
\nonumber \\
\langle P_{hT}^{2} \rangle_{{}_{\rm BM}} & = & \langle P_{\perp}^{2} \rangle_{{}_{\rm C}} + z_{h}^{2}\,\langle k_{\perp}^{2} \rangle_{{}_{\rm BM}} .
\label{eq:eqn_FUU_PT_final}
\eea
Eqs.~(\ref{eq:eqn_FUU_TMD_final})-(\ref{eq:eqn_FUU_BM2_final}) and the other related formulas are used for calculating the central values of the unpolarized cross-section 
pseudo-data with and without the azimuthal modulations in 5D binning of \xbj, $Q^2, P_{hT}, z_{h}$ and  $\phi_{h}$ (see Appendix \ref{sec:appB} as well as Appendices A and B 
in \cite{RG_E12-10-006_12-11-007}). The pseudo-data are obtained from an event generator package, the original version of which can be found at \cite{Liu:2019}. In the analysis, 
we take
\bea
\langle k_{\perp}^{2} \rangle & = & 0.604~{\rm (GeV/c)^{2}} \,,
\nonumber \\
\langle P_{\perp}^{2} \rangle & = & 0.114~{\rm (GeV/c)^{2}} \,,
\label{eq:eqn_kT_PT}
\eea
along with all the parameters in
Eqs.~(\ref{eq:eqn_FUU_TMD_final})-(\ref{eq:eqn_FUU_BM2_final}) taken from \cite{Bastami:2018xqd,Liu:2019,Liu:2022}: namely,
\begin{itemize}
\item[$\bullet$] $N_{u} =0.35$, $N_{d} = -0.9$, $\alpha_{u} = 0.73$, $\beta_{u} = 3.46$, 
$\alpha_{d} = 1.08$, $\beta_{d} = 3.46$, $M_{{\!}_{\rm BM}}^{2} = 0.34~{\rm GeV^{2}}$;

\item[$\bullet$] $N_{u,fav}^{\rm C} = 0.49$, $N_{d,dis}^{\rm C} = -1.00$, $\gamma = 1.06$, $\delta = 0.07$, 
$M_{\rm C}^{2} = 1.50~{\rm GeV^{2}}$.
\end{itemize}
Notice that the choice in Eq.~(\ref{eq:eqn_kT_PT}) is somewhat consistent with the findings in \texttt{MAPTMD24} (see the mean values of flavor-independent pink points in Fig.~19 of Ref.~\cite{Bacchetta:2024qre}) and similar to Set V of parameters discussed in Ref.~\cite{Christova:2020ahe}.

We also note that while in the structure function $F_{UU}$, the Gaussian widths for the unpolarized TMD PDFs and TMD FFs enter only in a specific combination (see 
Eq.~(\ref{eq:eqn_FUU_PT_final})), in both Cahn contributions, Eqs.~(\ref{eq:eqn_FUU_Cahn_final}) and (\ref{eq:eqn_FUU_Cahn2_final}), $\langle k_{\perp}^{2} \rangle$ appears 
explicitly. This means that more precise data in specific/extended kinematical regions will offer an important and complementary tool for the study of these properties.

\section{Nuclear effects from Fermi motion}
\label{sec:nucl_corr}

\begin{figure}[tbh!]
\hspace{-1.25cm}
   \begin{minipage}{0.425\textwidth}
       \includegraphics[width = 1.125\textwidth]{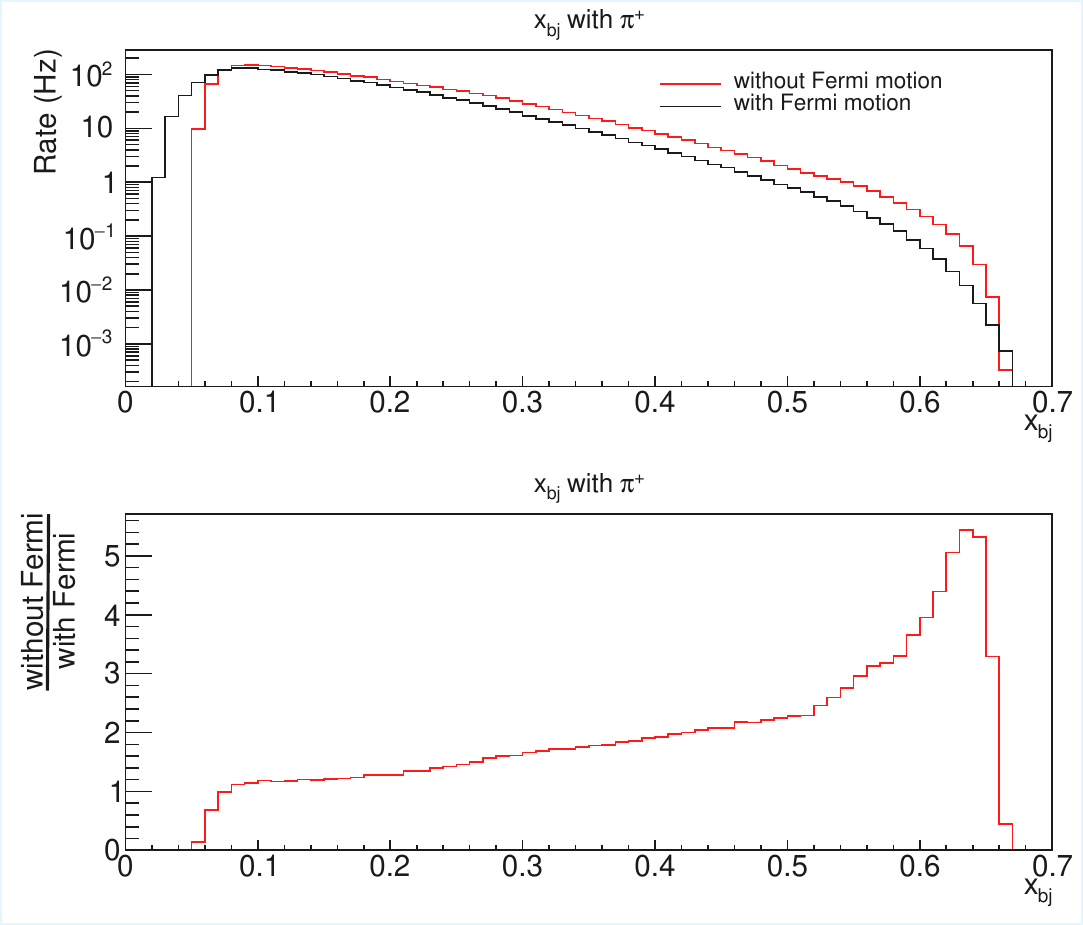}
   \end{minipage}
   \caption{
   Upper panel: $x_{bj}$ distributions of generated events' rate for $\pi^{+}$ as leading hadron, with (black histogram) and without (red histogram) Fermi motion of the initial 
   nucleon in the $^{3}$He target. Lower panel: their ratio.}
   \label{fig:Q2_fermi}
\end{figure}

\begin{figure}[tbh!]
   \hspace{-1.25cm}
   \begin{minipage}{0.4\textwidth}
       \includegraphics[width = 1.15\textwidth]{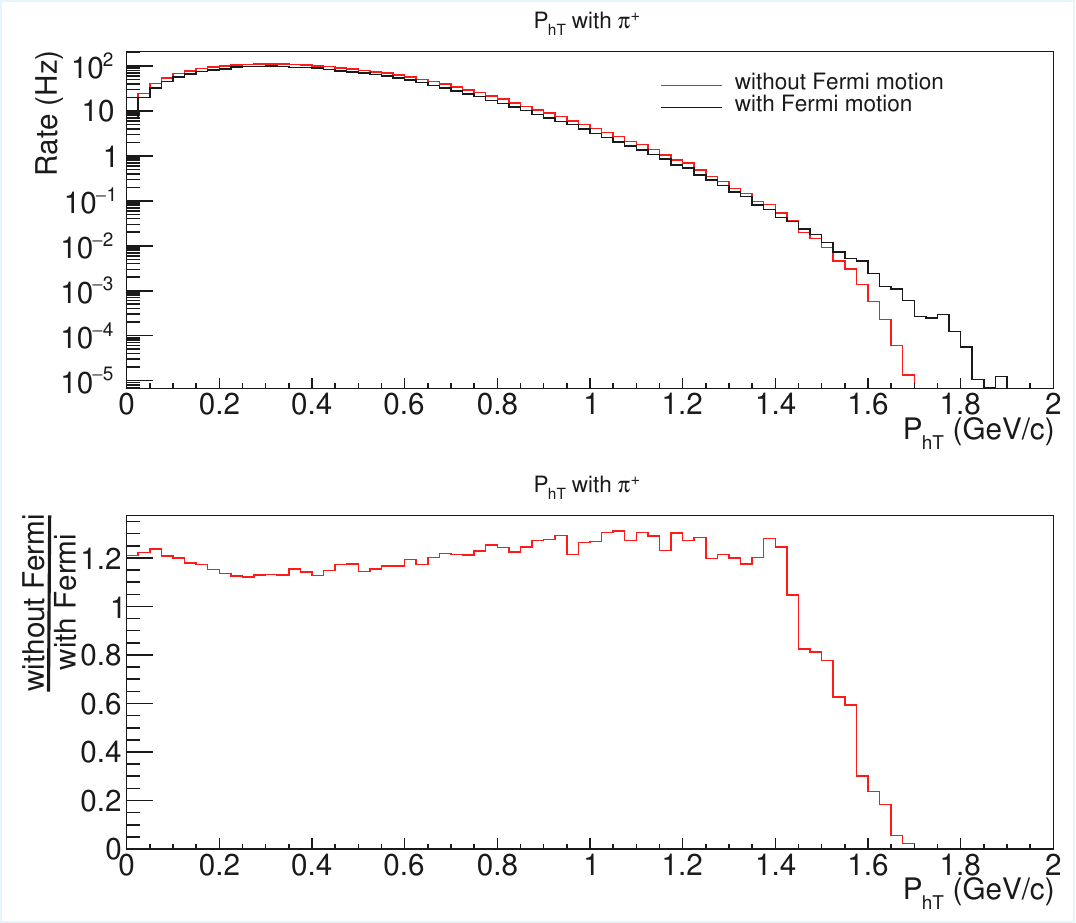}
   \end{minipage}
   \caption{
   The first plot displays the $P_{hT}$ distributions of generated events' rate for $\pi^{+}$ as leading hadron, with (black histogram) and without (red histogram) 
   Fermi motion of the initial nucleon in the $^{3}$He target, whereas the second plot shows their ratio.}
   \label{fig:Pt_fermi}
\end{figure}

\begin{figure}[tbh!]
   \hspace{-1.25cm}
   \begin{minipage}{0.4\textwidth}
       \includegraphics[width = 1.15\textwidth]{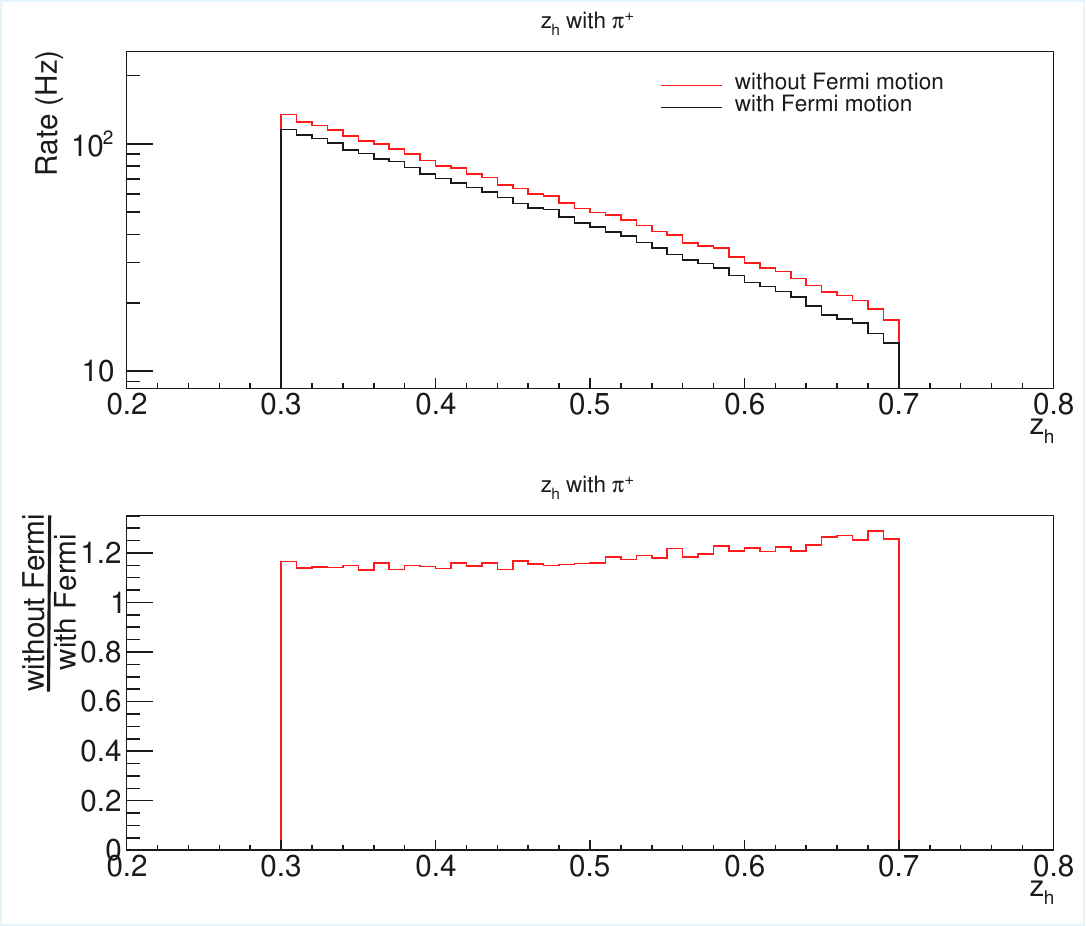}
   \end{minipage}
   \caption{
   The first plot displays the $z_{h}$ distributions of generated events' rate for $\pi^{+}$ as leading hadron, with (black histogram) and without (red histogram) 
   Fermi motion of the initial nucleon in the $^{3}$He target, whereas the second plot shows their ratio.}
   \label{fig:Pt_fermi_z}
\end{figure}

We consider here possible sources of nuclear effects to estimate their effect on the process under investigation. Indeed, in the $^{3}$He rest frame the initial-state nucleon 
entering the SIDIS process in Eq.~\eqref{eq:eqn_SIDIS} is not at rest due to its Fermi motion. 
This intrinsic motion modifies the kinematics relative to the scattering off a stationary neutron (or proton) in the laboratory frame, shifting, in particular, the variables \xbj, 
$P_{hT}$, and $z_{h}$. Consequently, this can affect the extraction of the transverse momentum widths $\langle k_{\perp}^2 \rangle$ and $\langle P_{\perp}^2 \rangle$
(see e.g., Eq.~\ref{eq:eqn_kT_PT}). Within the kinematic ranges considered here and relevant to the proposal in Ref.~\cite{RG_E12-10-006_12-11-007}, we have found that the differences 
in \xbj, $P_{hT}$, and $z_{h}$ due to the Fermi motion can lead to average changes of approximately $20\%$, as illustrated by ratio plots in 
Figs.~\ref{fig:Q2_fermi}–\ref{fig:Pt_fermi_z}\footnote{These results refer to the production of positively charged pions at 11~GeV electron beam energy.}. 

\begin{figure}[h!]
\hspace{-1.5cm}
   \begin{minipage}{0.4\textwidth}
       \includegraphics[width = 1.2\textwidth, height=0.7\textwidth]{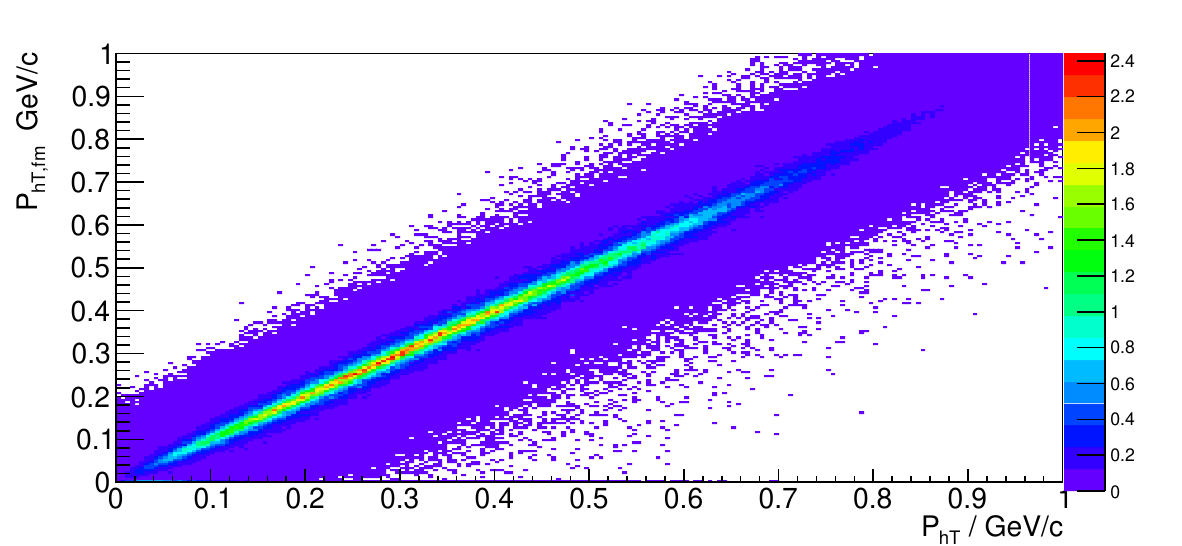}
   \end{minipage}
   \caption{
   A 2D plot showing the $P_{hT}$ distribution with Fermi motion versus without Fermi motion of the initial nucleon in the $^{3}$He target for $\pi^{+}$ as the leading hadron, 
   weighted by the cross section $\times$ acceptance.}
   \label{fig:Pt_fermi_2D}
\end{figure}

\begin{figure}[h!]
\hspace{-0.45cm}
    \includegraphics[width=1.0\linewidth]{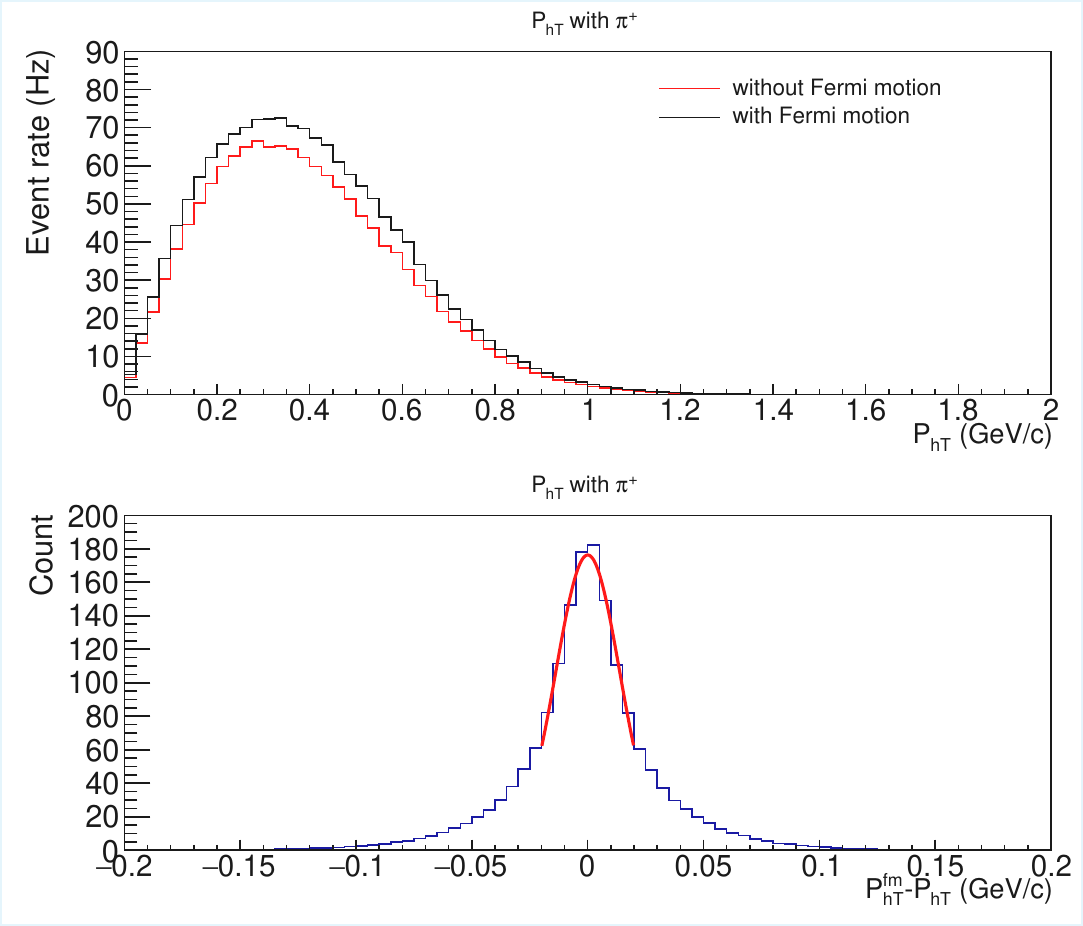}
    \caption{
    Upper panel: $P_{hT}$ distribution with and without Fermi motion of the initial nucleon in the $^{3}$He target for $\pi^{+}$ as leading hadron. Lower panel: their difference.}
    \label{fig:Pt_diff}
\end{figure}

In particular, modifications in $P_{hT}$ directly impact the extracted transverse momentum widths, according to Eq.~\eqref{eq:eqn_mom_cons}. In this case, the obtained correlation between $P_{hT}$ 
values with and without Fermi motion is shown in Fig.~\ref{fig:Pt_fermi_2D} (see also Fig.~\ref{fig:Pt_diff}). Furthermore, the comparison of the $P_{hT}$ distributions with and without Fermi motion 
-- using $\pi^{\pm}$ as the leading hadron -- shows that the smearing effect on $P_{hT}$ is approximately $\pm 0.014~{\rm GeV/c}$. As shown in \cite{RG_E12-10-006_12-11-007}, one can see that the
results of Figs.~\ref{fig:Q2_fermi}, \ref{fig:Pt_fermi}, \ref{fig:Pt_fermi_2D}, and \ref{fig:Pt_diff} are quite similar to those of $\pi^{-}$ as leading hadron. This smearing 
translates to uncertainties of $\pm 0.0006~{\rm (GeV/c)^{2}}$ in the extracted $\langle k_{\perp}^2 \rangle$, as well as $\pm 0.0001~{\rm (GeV/c)^{2}}$ in $\langle P_{\perp}^2 \rangle$. 
These results indicate that while the Fermi motion introduces non-negligible effects, those can be properly quantified and systematically studied. In the proposed data analyses of the future SoLID 
SIDIS experiments E12-10-006 and E12-11-007, appropriate correction procedures to account for the Fermi motion effects will be implemented, ensuring accurate extraction of the intrinsic transverse 
momentum distributions. Nevertheless, the impact of the nuclear effects based on the Fermi motion (within the current ansatz) on the statistical uncertainty part of the SoLID pseudo-data has been 
studied and found to be negligible as compared to the total uncertainties or, even the systematic uncertainties (see Appendix~\ref{sec:appA}), and therefore does not affect the physics impact results 
shown in Sec.~\ref{sec:res}.

We also note that in the study of neutron single spin asymmetries from SIDIS off a transversely polarized $^3$He target at the SoLID SIDIS kinematics~\cite{Scopetta_2007}, it was shown 
that the Fermi motion and binding effects can be properly taken into account via a simple procedure: more precisely, the only nuclear structure ingredients turn out to be the nucleon 
effective polarizations (quantities known from precise few-body calculations in a rather model independent way). 

Furthermore, more recently, Liu and collaborators~\cite{Liu:2022}, focusing on unpolarized SIDIS off light nuclei, have shown that the conventional nuclear effect on the unpolarized 
structure function can be at a few percent level and even smaller for the azimuthal asymmetry.  These results may provide a baseline for new studies of other exotic nuclear effects and 
serve as an estimate for nuclear corrections in experimental analyses of future data involving light nuclear targets (such as the proposed SoLID SIDIS data analyses of \cite{E12-10-006},
\cite{E12-11-007}, and \cite{RG_E12-10-006_12-11-007}). 

The approved run-group proposal~\cite{RG_E12-10-006_12-11-007} and this paper may motivate more theoretical studies of nuclear effects in SIDIS processes with a $^3$He target. 
Together with the data, these new studies will not only enable reliable extractions of the information on the neutron, but also further investigation of possible EMC effect 
\cite{EMC1,EMC2,JeffersonLabHallATritium:2024las} in TMD physics.

\section{Physics impact studies}
\label{sec:res}
In this section we present our projections and \texttt{MAPTMD24+SoLID} physics impact results on the unpolarized cross section, including the SoLID statistical and systematic uncertainties 
(see Appendix~\ref{sec:appA} and Ref.~\cite{RG_E12-10-006_12-11-007}).

Adopting the \texttt{MAPTMD24} framework~\cite{Bacchetta:2024qre}, as elaborated in Sec.~\ref{sec:phi-integrated}, we show the impact of the SoLID pseudo-data on the unpolarized TMDs 
(extracted in the analysis of the MAP Collaboration) in Sec.~\ref{sec:unpol_res2} in particular.

\subsection{SoLID SIDIS kinematics}
\label{sec:unpol_res0}

In our study we have used the 
following kinematic-variable ranges of the pseudo-data sets given as  
\begin{itemize}
\item[] \!\!\!\!\!\!\!$0 <$ \xbj $< 0.7$, ~~$0.3 < z_{h} < 0.7$, ~~$Q^{2} > 1.4~{\rm GeV^{2}}$,
\item[] \!\!\!\!\!\!\!$P_{hT} < \mbox{min}\left[\mbox{min}[0.2\,Q, ~0.5\,z_{h}\,Q] + 0.3~{\rm GeV}, ~z_{h}\,Q \right]$.
\end{itemize}

\begin{figure}[hbt!]
\centering
\hspace{0.0cm}
\includegraphics[width=7.0cm]{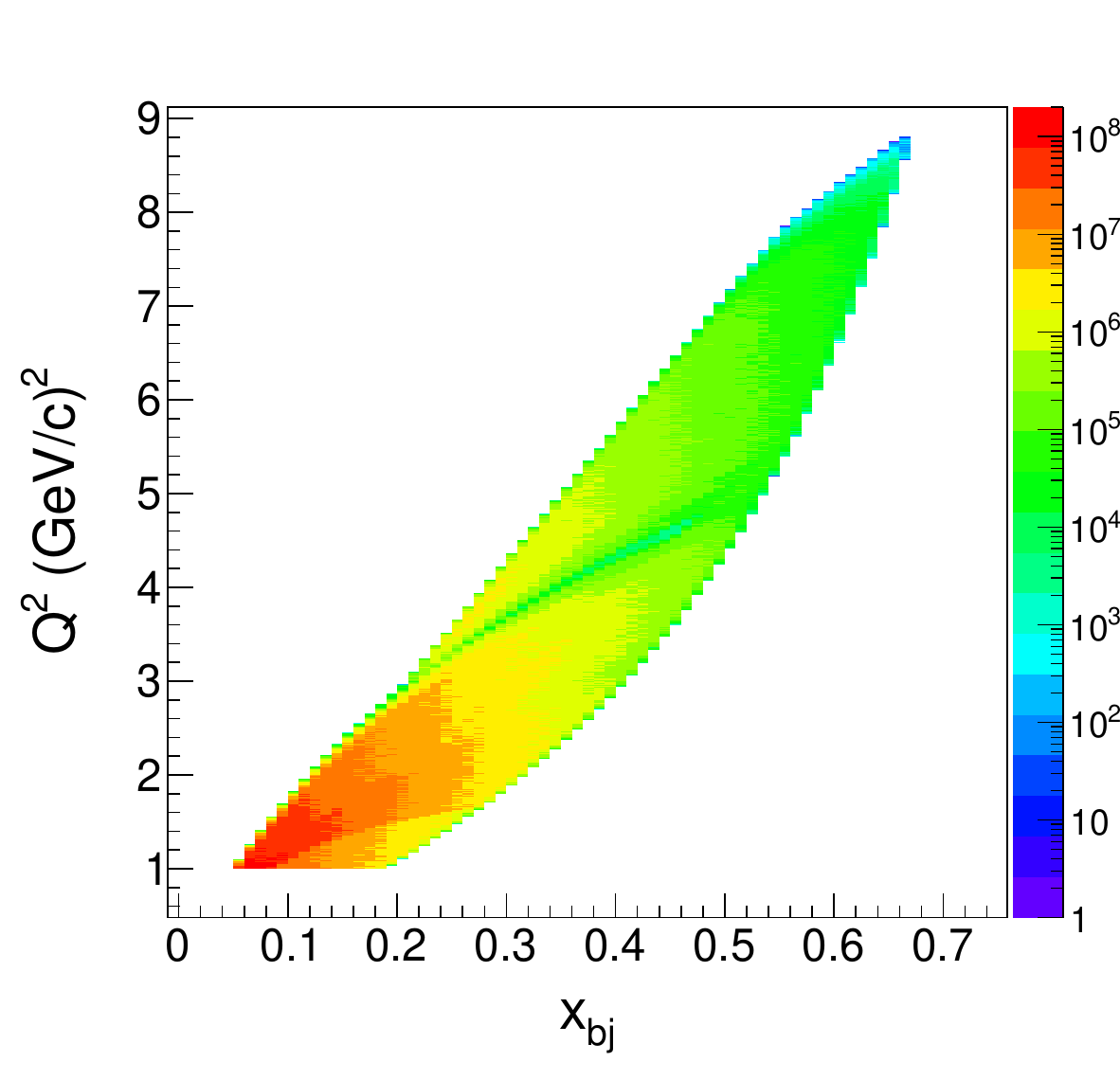}
\includegraphics[width=7.0cm]{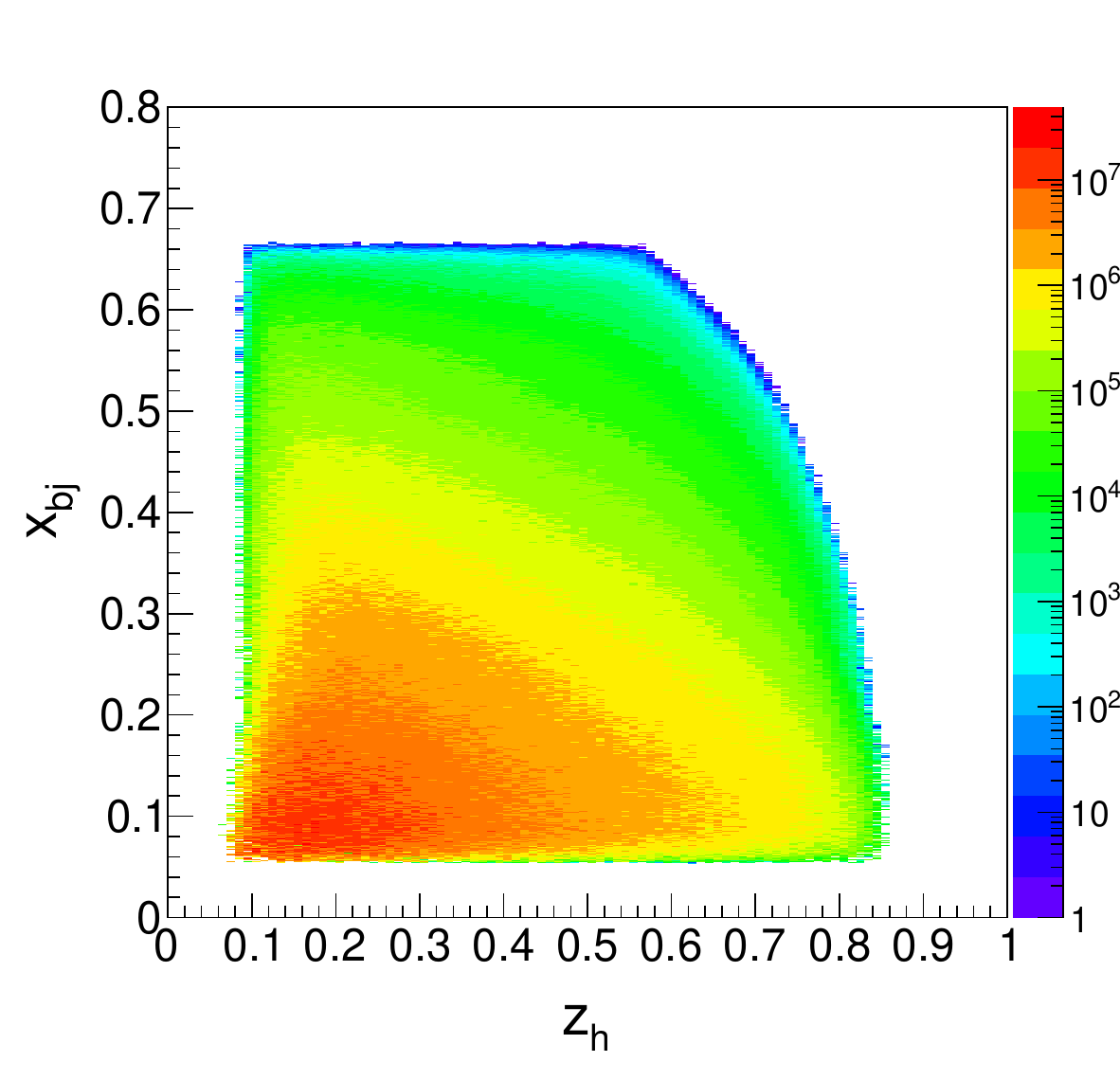}
\\[-0.1cm]
\caption{Phase-space correlations for the SoLID apparatus between $Q^{2}$ and \xbj  (top plot), \xbj and $z_{h}$ (bottom plot).
The SoLID $z_{h}$ cut is not applied.}
\label{fig:fig_ps}
\end{figure}

In Fig.~\ref{fig:fig_ps} we show the correlations between different kinematic variables (equivalently, the phase-space components) obtained taking into account the SoLID acceptance 
but without imposing the $z_{h}$ cut. More precisely, we present the correlations in the phase space at the combined 11~GeV and 8.8~GeV beam energies for $\pi^{+}$ production off 
transversely polarized ${}^{3}$He target. 

\subsection{Transverse momentum projections for the SIDIS unpolarized cross section}
\label{sec:unpol_res1}

\begin{figure}[hbt!]
\centering
    \includegraphics[width=8.75cm]{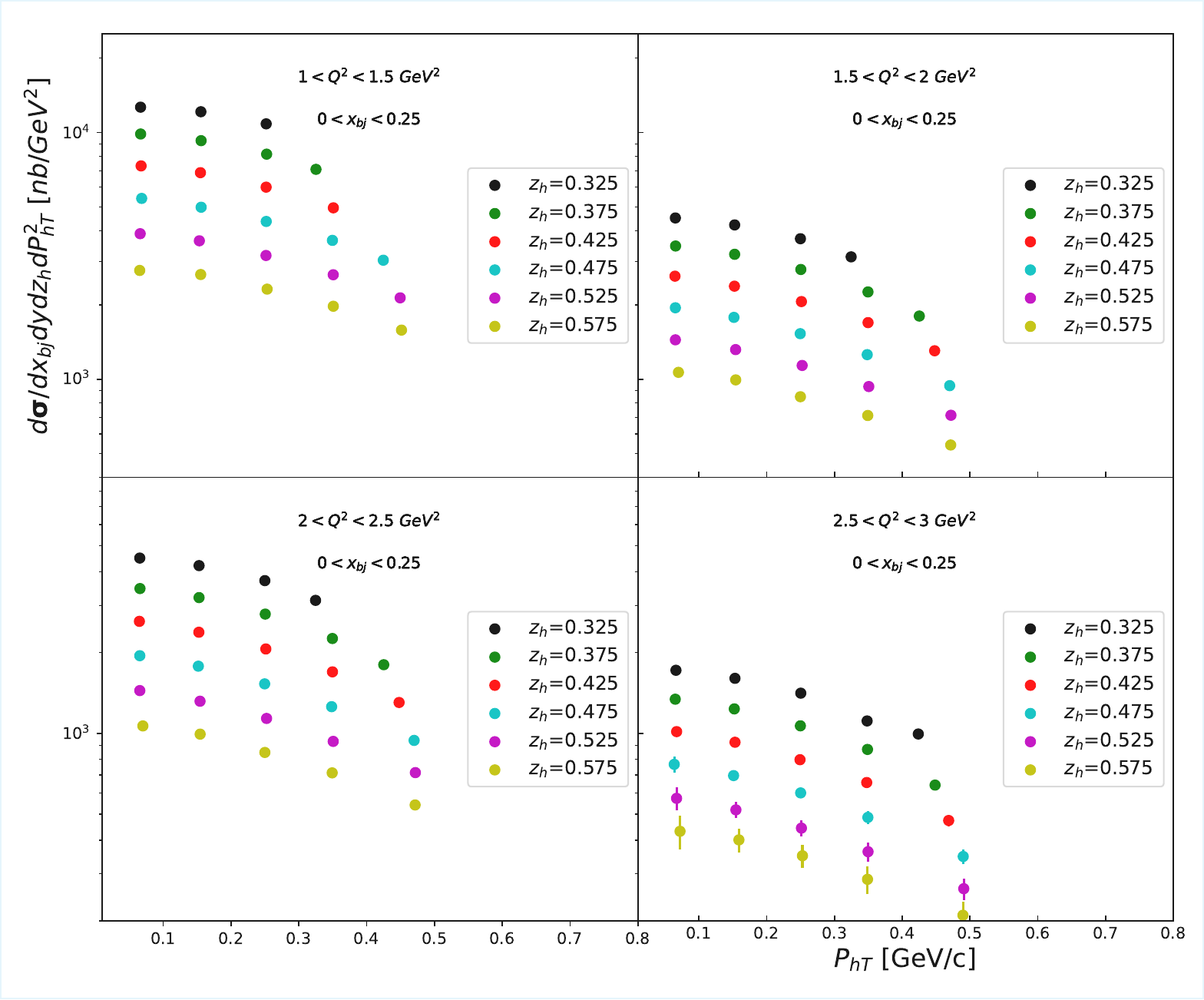}
    \vspace{-0.5cm}
\caption{Unpolarized cross section for $\pi^+$ production at beam energy 11~GeV as a function of $P_{hT}$, in specific \xbj and $Q^2$ kinematic bins, 
and at given $z_{h}$ values within the range of [0.3, 0.7].}
\label{fig:11_cs_pTa}
\end{figure}

\begin{figure}[hbt!]
\centering
    \includegraphics[width=8.75cm]{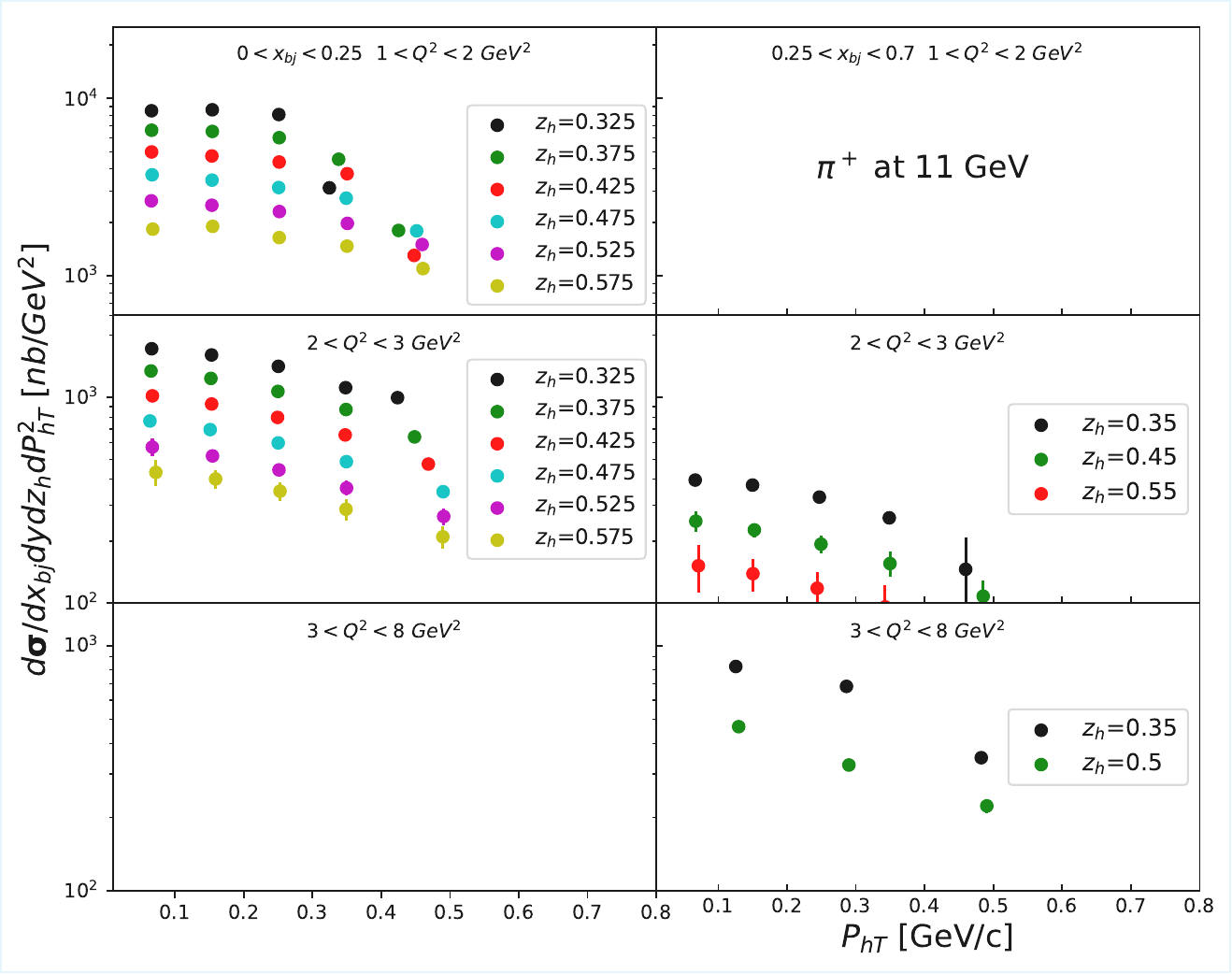}
    \vspace{-0.5cm}
\caption{Unpolarized cross section for $\pi^+$ production at beam energy 11~GeV as a function of $P_{hT}$, as in Fig.~\ref{fig:11_cs_pTa} but for 
larger two \xbj and three $Q^2$ kinematic bins, and at given $z_{h}$ values within the range of [0.3, 0.7]. }
\label{fig:11_cs_pTb}
\end{figure}

We start by considering the SoLID projections of the SIDIS unpolarized cross section (see Eq.~(\ref{eq:eqn_FUU})) integrated over $\phi_{h}$ as a function of $P_{hT}$ for $\pi^{+}$. 
Examples of such projections are shown in Figs.~\ref{fig:11_cs_pTa} and~\ref{fig:11_cs_pTb}.
The pseudo-data points in both figures include the SoLID statistical and systematic uncertainties combined in quadrature, and the central values are determined by employing the \texttt{MAPTMD24} framework (more details are collected in  Sec.~\ref{sec:unpol_res2}). In all the panels of both figures, the pseudo-data are simulated at 11~GeV. 
In addition, in Appendix~\ref{sec:appB}, where the central values of the projections are obtained using the general ansatz of a simpler TMD parton model discussed in Sec~\ref{sec:phi-dependent}, 
we show results both at 11~GeV and 8.8~GeV.

\subsection{Impact results on unpolarized TMDs and FFs}
\label{sec:unpol_res2}

\begin{figure}[h!]
\hskip -0.25truecm
\includegraphics[width=1.0\linewidth]{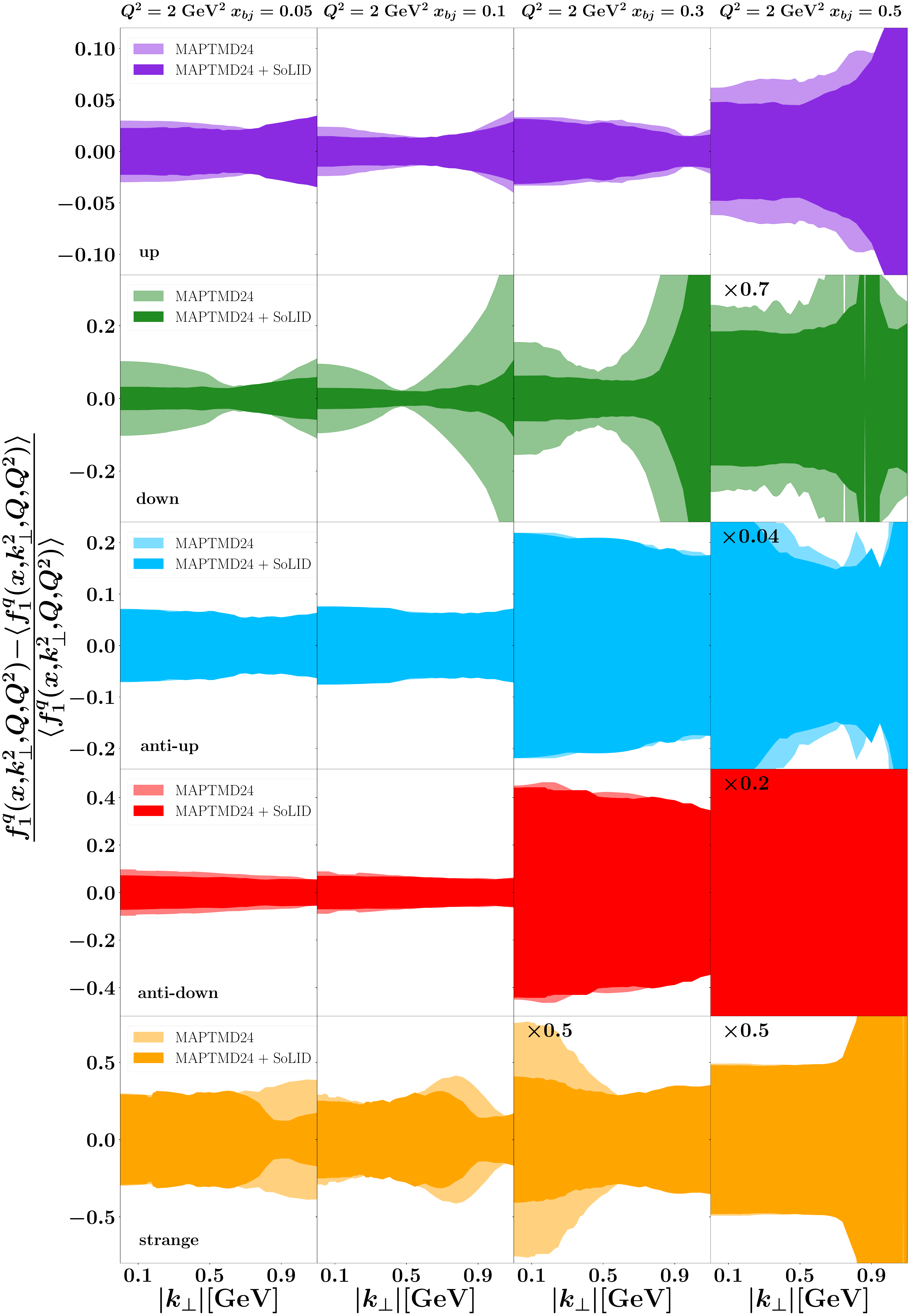}
\vskip 0.25truecm
\caption{The \texttt{MAPTMD24} and impact \texttt{MAPTMD24+SoLID} results on unpolarized TMDs for the final-state pions \& kaons configuration in the bins 
of $x_{{\!}_{bj}} = 0.05$, $x_{{\!}_{bj}} = 0.1$, $x_{{\!}_{bj}} = 0.3$, $x_{{\!}_{bj}} = 0.5$ at $Q^{2} = 2~{\rm GeV^{2}}$ 
as a function of $|k_{\perp}|$. See the text in this section for more details.}
\label{fig:fig_pika_impact1}
\end{figure}

The unpolarized TMD-related impact results are shown in Fig.~\ref{fig:fig_pika_impact1}
(for $\pi^{\pm}$ and $K^{\pm}$), as well as in Appendix~\ref{sec:appC}, in Fig.~\ref{fig:fig_pika_impactC3}
(for $\pi^{\pm}$) and 
in Fig.~\ref{fig:fig_pika_impactC6}
(for $K^{\pm}$). All figures
are obtained at $Q^2=2$ ${\rm GeV^{2}}$. Qualitatively similar results are also obtained at other values of $Q^2$ available at SoLID: namely, for 4 and 6~${\rm GeV^{2}}$ \cite{RG_E12-10-006_12-11-007}. 
Each column refers to a fixed \xbj value: namely, 0.05, 0.1, 0.3 and 0.5, while each row refers to a given quark flavor in the \texttt{MAPTMD24} ansatz, where 
the flavors considered are $u$, $d$, $\bar{u}$, $\bar{d}$, and $s$. In the vertical axis we have the relative difference of the unpolarized TMD $f_{1}^{q}$ with respect to 
the average value ($\langle f_1^q\rangle$) of the ensemble of replicas from the \texttt{MAPTMD24} (light bands) and the \texttt{MAPTMD24+SoLID} (dark bands) analyses, so defined:
\beq
\frac{f_{1}^{q}(x_{{\!}_{bj}}, k_{\perp}^{2}, Q, Q^{2}) - \langle f_{1}^{q}(x_{{\!}_{bj}}, k_{\perp}^{2}, Q, Q^{2}) \rangle}
{\langle f_{1}^{q}(x_{{\!}_{bj}}, k_{\perp}^{2}, Q, Q^{2}) \rangle} .
\label{eq:eqn_impact_eq}
\eeq
The horizontal axis shows the modulus of the quark intrinsic transverse momentum $|\bm{k}_{\perp}|$. The light and dark uncertainty bands are computed at 68\% 
confidence level (CL).

One can see that the main impact, coming from the SoLID $^3$He pseudo-data, is on the $d$-quark distribution. This is because the experimental data used in the \texttt{MAPTMD24} 
analysis are for proton and deuteron targets and do not provide constraints in the SoLID kinematic region. In contrast, the $u$-quark TMD is already well constrained by the 
available data, except in the large-$x$ region (\xbj $\simeq$ 0.5). It is important to stress that the uncertainties obtained in our study, by using the SoLID pseudo-data of 
the $\pi^{\pm}$ \& $K^{\pm}$ configuration, could be similar or, slightly larger than the uncertainties obtained by using only the pseudo-data of $\pi^{\pm}$. This can be traced 
back to the following issues: (i) unavoidable intrinsic uncertainty stemming from the used collinear PDF sets; (ii) correlations coming from having similar TMDs but different 
cross sections in certain regions; (iii) specific number of replicas that is fixed by the \texttt{MAPTMD24} extraction.
\begin{figure}[h!]
\hskip -0.25truecm
\includegraphics[width=0.95\linewidth]{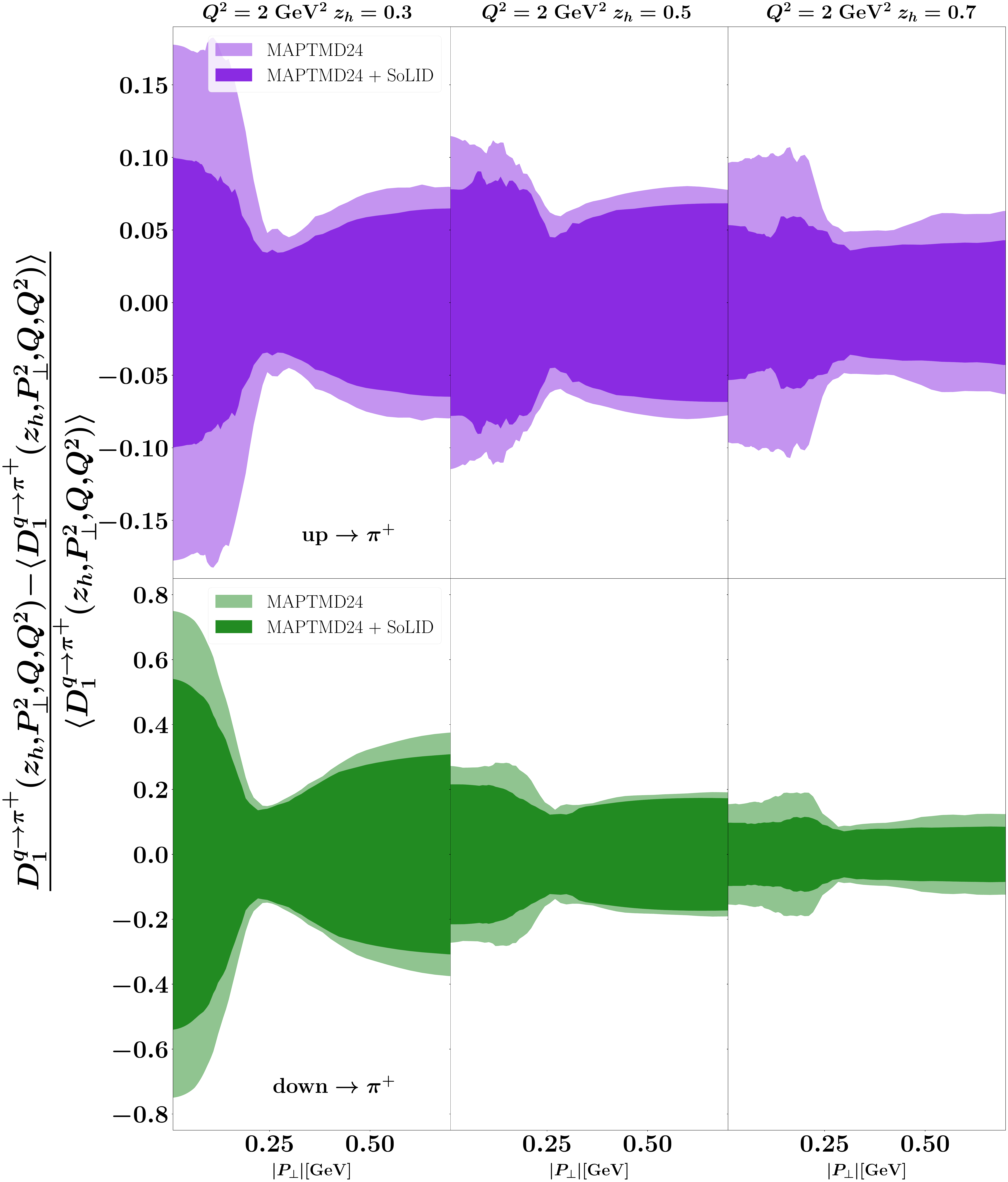}
\vskip 0.25truecm
\caption{The \texttt{MAPTMD24} and impact \texttt{MAPTMD24+SoLID} results on unpolarized FFs for the final-state pions in the bins 
of $z_{h} = 0.3$, $z_{h} = 0.5$, $z_{h} = 0.7$ at $Q^{2} = 2~{\rm GeV^{2}}$ as a function of $|P_{\perp}|$. See the text in this 
section for more details.}
\label{fig:fig_pi_impact1}
\end{figure}
\begin{figure}[h!]
\hskip -0.25truecm
\includegraphics[width=0.95\linewidth]{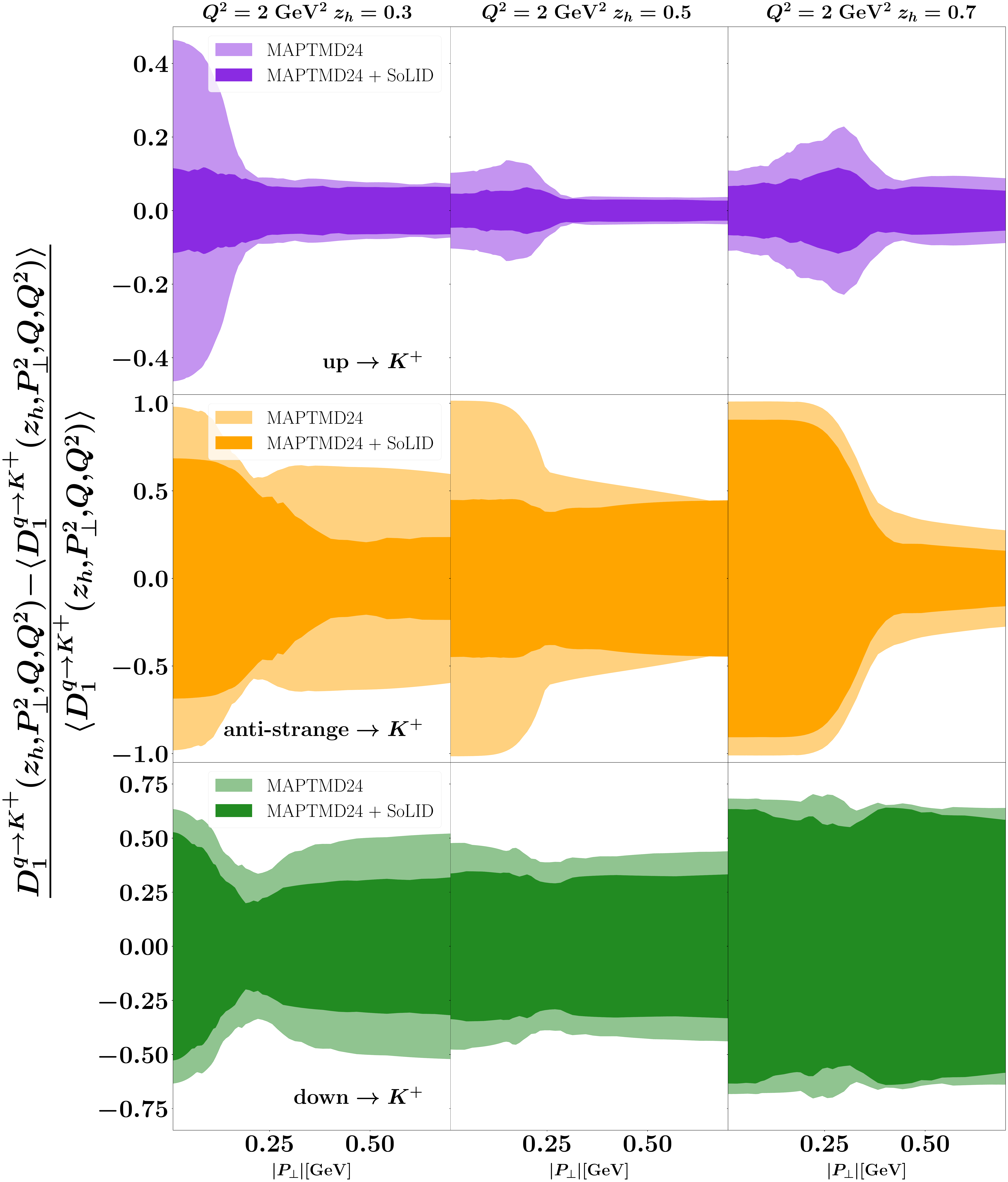}
\vskip 0.25truecm
\caption{The \texttt{MAPTMD24} and impact \texttt{MAPTMD24+SoLID} results on unpolarized FFs for the final-state kaons in the bins 
of $z_{h} = 0.3$, $z_{h} = 0.5$, $z_{h} = 0.7$ at $Q^{2} = 2~{\rm GeV^{2}}$ as a function of $|P_{\perp}|$. See the text in this 
section for more details.}
\label{fig:fig_ka_impact2}
\end{figure}

For completeness, in Figs.~\ref{fig:fig_pi_impact1} and \ref{fig:fig_ka_impact2} we show the impact study on the pion and kaon TMD FFs, the complementary nonperturbative key element in SIDIS. 
The quantity under consideration is the corresponding one appearing in Eq.~(\ref{eq:eqn_impact_eq}), which is
\beq
\frac{D_{1}^{q\to h}(z_{{\!}_{h}}, P_{\perp}^{2}, Q, Q^{2}) - \langle D_{1}^{q\to h}(z_{{\!}_{h}}, P_{\perp}^{2}, Q, Q^{2}) \rangle}
{\langle D_{1}^{q\to h}(z_{{\!}_{h}}, P_{\perp}^{2}, Q, Q^{2}) \rangle} .
\label{eq:eqn_impactFF_eq}
\eeq
More in detail, in Fig.~\ref{fig:fig_pi_impact1} we present the results for the favored ($u\to \pi^+$) and disfavored ($d\to\pi^+$) pion TMD FFs\footnote{Here favored (disfavored) refers 
to the case where the hadron contains (does not contain) the fragmenting quark as a valence quark.} at fixed scale, $Q^2=2$ GeV$^2$, for different values of $z_h$ as a function of 
$|\bm{P}_\perp|$. Similarly, in Fig.~\ref{fig:fig_ka_impact2} we show the results for the three relevant kaon TMD FFs, namely $u\to K^+$, $\bar s\to K^+$ (both favored) and $d\to K^+$ 
(disfavored) for the same $Q^2$ and $z_h$ values. 
As one can see even in the case of TMD FFs the reduction in the uncertainties is significant, showing once again the impact of the SoLID analysis.


\subsection{TMD width parameters}
\label{sec:unpol_res3}

\subsubsection{Results from the parton model analysis of azimuthal modulations}
\label{sec:unpol_res3b}



\begin{figure}[h!]
\vspace{0.0cm}
\includegraphics[width=1.015\linewidth]{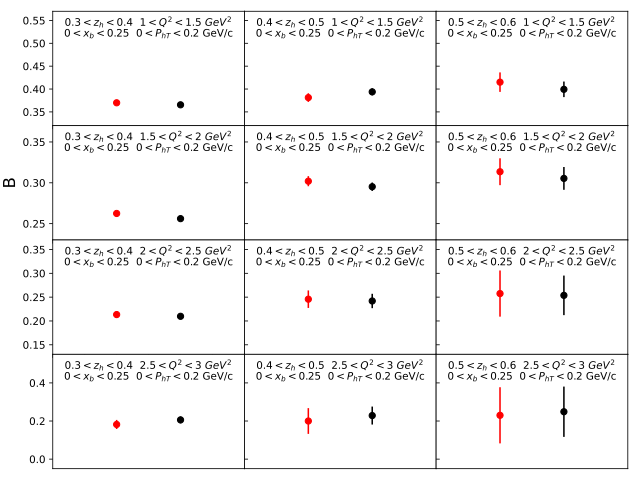}
\vspace{-0.5cm}
    \caption{The parameter B at fixed \xbj and $P_{hT}$ bins for different $z$ and $Q^2$ bins (see the legend) as obtained by using the fitting functional form of 
    $\mbox{A}(1 - \mbox{B}\cdot\cos(\phi_h) - \mbox{C}\cdot\cos(2\phi_h))$. The red and black circle points represent the results 
    for $\pi^+$ and $\pi^-$ production channels, respectively.
    }
\label{fig:Bs}
\end{figure}

\begin{figure}[h!]
\vspace{0.0cm}
\includegraphics[width=1.015\linewidth]{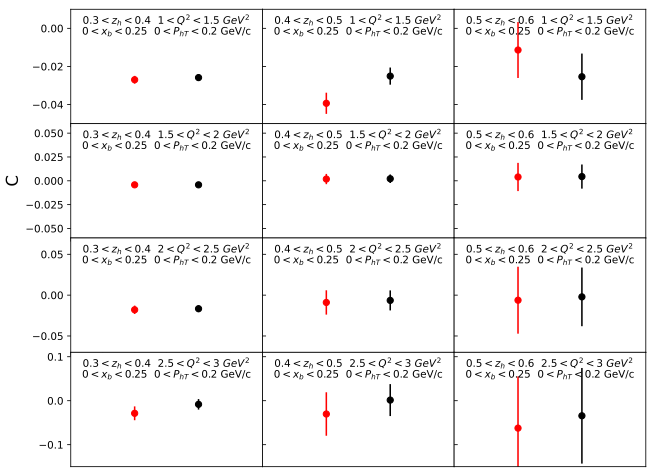}
\vspace{-0.5cm}
    \caption{The parameter C at fixed \xbj and $P_{hT}$ bins for different $z$ and $Q^2$ bins (see the legend) as obtained by using the fitting functional form of 
    $\mbox{A}(1 - \mbox{B}\cdot\cos(\phi_h) - \mbox{C}\cdot\cos(2\phi_h))$. The red and black circle points represent the results 
    for $\pi^+$ and $\pi^-$ production channels, respectively.
    }
\label{fig:Cs}
\end{figure}

In this section we consider the SoLID pseudo-data for the $\phi_h$ distributions and try to fit them with a simple function, $\mbox{A}(1 - \mbox{B}\cdot\cos(\phi_h) - \mbox{C}\cdot\cos(2\phi_h))$, 
to evaluate the azimuthal modulation effects. In this formula, the parameters B and C indicate the amplitudes of the modulation  and are directly related to the structure functions entering 
Eq.~(\ref{eq:eqn_FUU}), while the parameter A is related to $F_{UU}$. By applying the corresponding simplified fitting, e.g., to the $\phi_h$-dependent $\pi^{+}$ pseudo-data from 
Fig.~\ref{fig:fig_xs_pip1} (and $\pi^{-}$ pseudo-data from Fig.~A9 in Appendix~A of \cite{RG_E12-10-006_12-11-007}) we obtain the results shown in Figs.~\ref{fig:Bs} and \ref{fig:Cs}.


These azimuthal modulation effects can be analyzed within a simplified TMD approach (Eqs.~(\ref{eq:eqn_FUU_TMD}), (\ref{eq:eqn_FUU_cos})
-(\ref{eq:eqn_FUU_BM}), (\ref{eq:eqn_FUU_cos2})-(\ref{eq:eqn_FUU_BM2}));
then, by adopting the Gaussian ansatz of Sec.~\ref{sec:phi-dependent}, the intrinsic transverse momentum widths 
can be extracted from the relative magnitudes of the structure functions $F_{UU}$, $F^{\cos(\phi_h)}_{UU}$ and $F^{\cos(2\phi_h)}_{UU}$ (Eqs.~(\ref{eq:eqn_FUU_TMD_final})-(\ref{eq:eqn_FUU_PT_final})). Notice that 
the two structure functions controlling the azimuthal modulations involve two contributions: the Cahn effect, Eqs.~(\ref{eq:eqn_FUU_Cahn}) and (\ref{eq:eqn_FUU_Cahn2}), 
and the less dominant Boer-Mulders effect, Eqs.~(\ref{eq:eqn_FUU_BM}) and (\ref{eq:eqn_FUU_BM2}). 
 
As one can see from Fig.~\ref{fig:Bs}, the parameter B is sizable (around 20-30\%) and, at least, for the lower and intermediate $z$ bins (left and central panels), with very small errors. 
Taking into account the explicit role of the $\langle k_\perp^2\rangle$ width in the Cahn effect (Eq.~(\ref{eq:eqn_FUU_Cahn_final})) entering this term, this quantity could be extremely 
useful and effective in the extraction of the Gaussian widths.

\begin{figure}[hbt!]
\centering
\hspace{0.0cm}
\includegraphics[width=7.5cm]{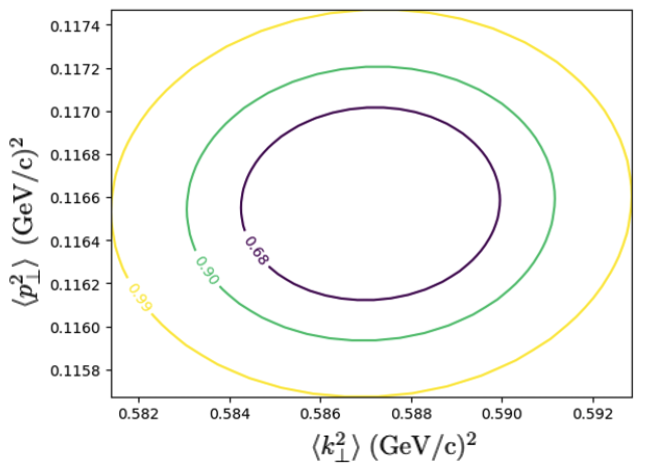}
    \caption{Three fitting contours for the Gaussian intrinsic width parameters, $\langle k_{\perp}^{2} \rangle$ and $\langle P_{\perp}^{2} \rangle$, obtained 
    for pion production using the functional form of the total SIDIS unpolarized cross section in Eq.~(\ref{eq:eqn_FUU}). Both Cahn and Boer-Mulders contributions 
    are included.}
\label{fig:kT_pT_pions}
\end{figure}

By substituting the expressions of the final (convoluted) structure functions from Eqs.~(\ref{eq:eqn_FUU_TMD_final})-(\ref{eq:eqn_FUU_PT_final}) into the SIDIS cross section in 
Eq.~(\ref{eq:eqn_FUU}), we obtain a fitting function in terms of the two intrinsic width parameters, $\langle k_{\perp}^2 \rangle$ and $\langle P_{\perp}^2 \rangle$. In order to 
carry out the projections with this fitting function we also use the (initial) values from Eq.~(\ref{eq:eqn_kT_PT}). The fitting is performed for pion production with all the available 
SoLID pseudo-data (employed to make all the figures in Appendices~A and B of \cite{RG_E12-10-006_12-11-007}), including the systematic uncertainties and using the Least Squares method.
The final results, presented by three contours with confidence levels of $68\%,90\%$ and $99\%$ are shown in Fig.~\ref{fig:kT_pT_pions}. The central values and their corresponding 
uncertainties for the 68\% CL are determined as
\bea
& & \langle k_{\perp}^{2} \rangle = 0.5868 \pm 0.0015~{\rm (GeV/c)^{2}} ,
\nonumber \\
& & \langle P_{\perp}^{2} \rangle = 0.1166\pm 0.0002~{\rm (GeV/c)^{2}} .
\label{eq:eqn_kT_PT2}
\eea

It is relevant to note that irrespective of the initial values used for the width parameters (which could be different from those adopted here), the obtained 
approximate result shows how one can significantly improve the accuracy in extracting this information. Thus, by measuring the unpolarized cross section with 
and without the azimuthal modulations, we will be able to extract the Gaussian width parameters $\langle k_{\perp}^{2} \rangle$ and $\langle P_{\perp}^{2} \rangle$. 
The extraction can therefore be carried out by fitting these two parameters to future data (using the functional forms given by Eqs.~(\ref{eq:eqn_FUU_TMD_final})-(\ref{eq:eqn_FUU_PT_final}) 
or, by new updated formulas within an improved TMD framework).

\subsubsection{Results from the NLO MAP analysis of the unpolarized cross sections}
\label{sec:unpol_res3a}

\begin{figure*}[hbt]
\centering
\hspace{-0.1cm}
\includegraphics[width=0.95\textwidth]{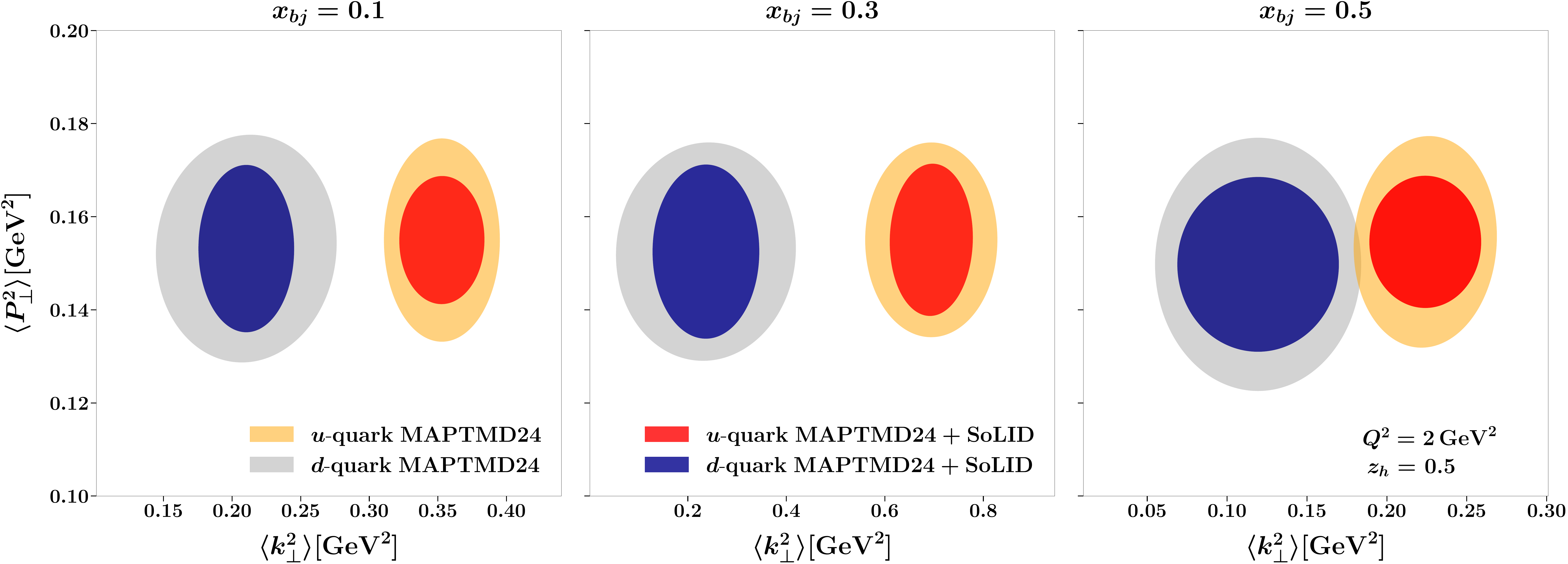} 
\caption{The contour bands at 68\% CL of $\langle \bm{P}^2_{\perp} \rangle^{q \rightarrow \pi^{+}}_{r}$ as a function of $\langle \bm{k}^2_{\perp} \rangle^{q}_{r}$, obtained from the fit 
of \texttt{MAPTMD24} and joint fit of \texttt{MAPTMD24+SoLID} for the $u$ (purple curves) and $d$ (green curves) quark flavors at fixed $z_{h} = 0.5$ and $Q^{2} = 2~{\rm GeV^{2}}$. The three panels correspond to three values of  \xbj: 
$x_{{\!}_{bj}} = 0.1$ (left panel), $x_{{\!}_{bj}} = 0.3$ (central panel) and $x_{{\!}_{bj}} = 0.5$ (right panel). 
}
\label{fig:fig_scatter_plot1}
\end{figure*}

We now move to a more sophisticated analysis focusing on the unpolarized cross sections within a proper TMD framework.
We will then  estimate the impact of the SoLID on the extraction of the widths of the distributions using the prescription from Ref.~\cite{Bacchetta2022} of MAP collaboration:
\bea
\!\!\!\!\!
\langle \bm{k}^2_{\perp} \rangle^{q}_{r}(x_{{\!}_{bj}},Q) =
\frac{2M_{N}^{2}\,\hat{f}_{1}^{q(1)}(x_{{\!}_{bj}},\bm{b}^2_{T},Q,Q^{2})} {\hat{f}_{1}^{q}(x_{{\!}_{bj}},\bm{b}^2_{T},Q,Q^{2})}\Bigg|_{|\bm{b}_T| = 2\,b_{\rm max}} ,
\label{eq:aver_kperp2}
\eea
where the subscript $r$ in the l.h.s.~refers to the \textit{regularized} definition of the average squared momenta, where $\hat{f}_{1}^{q}$ is the Fourier transform of the TMD PDF, and $\hat{f}_{1}^{q(1)}$ is its first Fourier moment~\cite{Boer:2011xd}. Analogously, one could apply similar arguments 
to the \textit{regularized} average squared transverse momentum for a final hadron $h$ in the fragmentation of a quark $q$
\cite{Bacchetta2022,Bacchetta:2019qkv,Boer:2011xd,Boer:2014bya}:
\bea
\langle \bm{P}^2_{\perp} \rangle^{q \rightarrow h}_{r}(z_{h},Q) & = &
\nonumber \\
& & \!\!\!\!\!\!\!\!\!\!\!\!\!\!\!\!\!\!\!\!\!\!\!\!\!\!\!\!\!\!\!\!\!\!\!\!\!\!\!
\frac{2z_{h}^{2}M_{h}^{2}\,\hat{D}_{1}^{q \rightarrow h(1)}(z_{h},\bm{b}^2_{T},Q,Q^{2})} {\hat{D}_{1}^{q \rightarrow h}
(z_{h},\bm{b}^2_{T},Q,Q^{2})}\Bigg|_{|\bm{b}_T| = 2\,b_{\rm max}} ,
\label{eq:aver_Pperp}
\eea
where $\hat{D}_{1}^{q \rightarrow h}$ is the Fourier transform of the TMD FF defined in Eq.~(\ref{eq:FTdefFF}), and $\hat{D}_{1}^{q \rightarrow h(1)}$ is the first Bessel moment~\cite{Boer:2011xd}
of the TMD FF.

In Fig.~\ref{fig:fig_scatter_plot1} we present the contour bands at 68\% CL of the average squared transverse momenta for the unpolarized TMD PDF and the TMD FF for a $\pi^+$ at fixed $Q^2=2$~GeV$^2$ and $z_h=0.5$, for different \xbj values: \xbj = 0.1, 0.3 and 0.5. 
Namely, we show $\langle \bm{P}^2_{\perp} \rangle^{q \rightarrow \pi^{+}}_{r}$ versus 
$\langle \bm{k}^2_{\perp} \rangle^{q}_{r}$ for the $u$ and $d$ quark flavors. The results are 
displayed for different ensembles of replicas with and without SoLID.

As one can see the uncertainties are strongly reduced, in all regions. Notice that these results cannot be directly compared with those shown in the previous subsection, since here we have both the \xbj/$z_h$ as well as the flavor dependence. Another method of studies of average values of $\langle \bm{k}^2_{\perp} \rangle^{q}$ and $\langle \bm{P}^2_{\perp} \rangle^{q}$ was proposed in Ref.~\cite{delRio:2024vvq} and can be employed on the real data of for estimations of the impact.

Ultimately, the future experimental data will be important to disentangle those dependencies.

\section{Conclusion and discussion}
\label{sec:sum}

In this paper we have presented the impact study results for the unpolarized SIDIS cross-section off $^3$He target based on the approved proposal of 
Ref.~\cite{RG_E12-10-006_12-11-007}, where further details can be found. In particular, the general results include $\pi^{+}$ and $\pi^{-}$ pseudo-data with the SoLID 
statistical uncertainties and theoretical central points projected into selected bins of ($x_{{\!}_{bj}}$, $P_{hT}, z_{h}, Q^{2}, \phi_{h})$ within SoLID kinematics, 
along with the systematic uncertainties evaluated for the acceptance correction, detection and identification of charged particles, radiative corrections, 
luminosity determination, and other sources. Charged-kaon contributions are considered as well. The SoLID systematic uncertainties are determined and assigned to the 
unpolarized (azimuthal-integrated) cross-section pseudo-data, without taking into account the azimuthal modulations. On the other hand, the results for the cross 
section including the azimuthal modulation terms are computed taking into account only the SoLID statistical uncertainties.

For the azimuthal-integrated cross section we adopt the advanced \texttt{MAPTMD24} framework with TMD evolution up to N$^3$LL accuracy (see Sec.~\ref{sec:phi-integrated}), 
whereas for the azimuthal angular-dependent cross section we employ a simpler TMD model (Generalized Parton Model), in which the transverse momentum dependence of TMDs 
is factorized and modeled in terms of Gaussians (see Sec.~\ref{sec:phi-dependent}). Though the latter approach is perhaps too simple to describe the experimental 
data, it can still be helpful to assess the experimental precision. 

Summarizing, in the paper we have presented physics impact results from the combined \texttt{MAPTMD24+SoLID} ansatz on the unpolarized TMDs as a function of $|\bm{k}_\perp|$ 
(for the PDFs) and $|\bm{P}_\perp|$ (for the FFs, both for pions and kaons) extracted in the \texttt{MAPTMD24} framework (see Sec.~\ref{sec:unpol_res2} and 
Appendix~\ref{sec:appC}). Besides, we show results on the (Gaussian) width parameter extraction using the \texttt{MAPTMD24+SoLID} joint fit (see Sec.~\ref{sec:unpol_res3a}), 
as well as their approximate results 
within a simplified TMD ansatz. We also present projection results of the SoLID pseudo-data on the unpolarized cross section as a function of $P_{hT}$ in various bins, 
showing the SoLID statistical and systematic uncertainties and central values as obtained employing the \texttt{MAPTMD24} framework (see Sec.~\ref{sec:unpol_res1}). 
The total systematic uncertainties (Table~\ref{tab:sys_tab} and Table~\ref{tab:sys_tab_kaon}, Appendix~\ref{sec:appA}) are applied only to the azimuthal-integrated cross 
sections, whereas the statistical uncertainties are applied to both cases\footnote{In the future, when a more accurate model becomes available, we will address the total 
systematic uncertainty of the unpolarized azimuthal angular-dependent cross section (expected to be of the same order of magnitude as that of the azimuthal-integrated 
cross section).}. In addition, Appendix~\ref{sec:appB} shows several projection results on the unpolarized cross section as a function of $\phi_{h}$ in various bins showing 
the SoLID statistical and systematic uncertainties, as well as central values obtained in the simplified TMD ansatz.

We have also discussed some results from the study on the effect of Fermi motion on the nucleon inside ${}^{3}$He. It has been found that the impact of this effect on the statistical 
uncertainty part of the SoLID pseudo-data is negligible as compared to the total uncertainties. However, more studies related to nuclear effects is currently ongoing. It is 
also important to mention that our proposed experiment will play an important role for the global studies of the EMC effect \cite{EMC1,EMC2,JeffersonLabHallATritium:2024las}, 
which addresses the differences in the partonic structure of a free proton and a bound proton in a nucleus. There are several possible explanations for the EMC effect, 
including the non-negligible Fermi motion of the nucleon, short-range correlations, the off-shellness of the bound nucleon and nuclear medium modifications. Combined with data 
on proton cross section~\cite{Bacchetta:2024qre,JAM20,H_xs_JLab,H_xs_NNPDF,Bacchetta2022}, the neutron cross section can be extracted using the state-of the-art 
calculation~\cite{PhysRevC.96.065203} for treating the $^3$He nuclear effect. Both in our projections and \texttt{MAPTMD24+SoLID} physics impact results, we have used the impulse 
approximation, in which the $^3{\rm He}$ is treated as a system of two protons plus one neutron. The extracted neutron results can then be compared to available and future data 
on deuterium target as well as upcoming results from various global analyses, which will allow for an improved investigation of the EMC effect in SIDIS processes considering the 
scattering off a $^3$He target.

\acknowledgments 

The work of H.G., S.J., V.K., Z.Z. is supported by the U.S. Department of Energy contract No.~DE-FG02-03ER41231.
The work of M.C., A.P., J.-P.C. is supported by the U.S. Department of Energy contract No.~DE-AC05-06OR23177, under which Jefferson Science Associates, LLC operates Jefferson Lab, 
and the National Science Foundation under Grants No.~PHY-2310031, No.~PHY-2335114 (A.P.). 
The work of Y.T. is supported by the U.S. Department of Energy contract No.~DE-FG02-84ER40146.
The work of U.D. is supported by the European Union ‘‘Next Generation EU’’ program through the Italian PRIN 2022 grant n. 20225ZHA7W. 
The work of L.R. is partially supported by the Italian Ministero dell'Universit\`a e Ricerca (MUR) through the research grant 20229KEFAM (PRIN2022, Next Generation EU, CUP H53D23000980006).


\appendix
\renewcommand{\theequation}{A\arabic{equation}}
\setcounter{equation}{0}
\renewcommand{\thefigure}{A\arabic{figure}}
\renewcommand{\theHfigure}{A\arabic{figure}}
\setcounter{figure}{0}

\section{\label{sec:appA} \\
Systematic uncertainties of the unpolarized cross section with $\mathrm{\bf F_{UU}}$}

In this appendix we present a summary of all SoLID systematic uncertainty sources studied and discussed in Sec.~5 of the experimental proposal of Ref.~\cite{RG_E12-10-006_12-11-007}; the magnitudes of the assigned systematic uncertainties are listed in Table~\ref{tab:sys_tab} (for pions) and Table~\ref{tab:sys_tab_kaon} (for kaons).

The total systematic uncertainty of the unpolarized cross section for charged pion production is up to $11\%$, calculated by rounding off the quadrature sum of the individual contributions. 
The largest source of uncertainty comes from the so-called coincidence acceptance correction being $8.2\%$ (which includes the electron acceptance correction uncertainty). One should also note that the $\pi^{+}$ and $\pi^{-}$ systematic uncertainties are determined to be approximately similar by their magnitudes. It is the case for both $\pi^{+}$ and $\pi^{-}$ at 
low-$Q^{2}$ and high-$Q^{2}$ regions as well. It is the reason why we consider only one total systematic uncertainty for both particles and for the entire $Q^{2}$ region, from low $Q^2$ to high $Q^2$.
\begin{table}[h!]
    \centering
    \begin{tabular}{|c|c|}
         \hline
         Sources & Uncertainty  \\
         \hline
         Coincidence acceptance correction & $8.2\%$ \\
         \hline    
         Experimental resolution & $3.5\%$ \\
         \hline        
         Pion detection efficiency & $ 4\%$ \\
         \hline
         Electron  detection efficiency & $< 2\%$ \\
         \hline
         Radiative corrections & $2.1\%$ \\
         \hline
         Vector meson production & $1\%$ \\
         \hline
         Luminosity determination & $\lesssim 3\%$ \\       
         \hline
         \hline
         Total & $\lesssim 11\%$ \\
         \hline
    \end{tabular}
    \caption{Systematic uncertainty budget of the unpolarized cross section for the produced (final-state) charged pions.}
    \label{tab:sys_tab}
\end{table}

It is essential to emphasize that for the unpolarized cross section for charged kaons all the systematic uncertainty sources, except for the coincidence acceptance correction and the charged-hadron detection efficiency, is the same as shown in Table~\ref{tab:sys_tab} for charged pion production. Meanwhile, for the kaons, the systematic uncertainty of the
\begin{itemize}
\item[(i)] coincidence acceptance correction is evaluated in a similar way as the acceptance correction for the pions discussed in Sec.~5.1.3 of \cite{RG_E12-10-006_12-11-007}. 

\item[(ii)] detection efficiency is evaluated in a similar way as the efficiency for the pions discussed in Sec.~5.3.1 of \cite{RG_E12-10-006_12-11-007}.
\end{itemize}
Thereby, the total systematic uncertainty of the unpolarized cross section for produced charged kaons is determined to be up to $18\%$, calculated by the quadrature sum 
of the individual contributions as shown in Table~\ref{tab:sys_tab_kaon}.
\begin{table}[h!]
    \centering
    \begin{tabular}{|c|c|}
         \hline
         Sources & Uncertainty  \\
         \hline
         Coincidence acceptance correction & $\sim 13\%$ \\ 
         \hline
         Experimental resolution & $3.5\%$ \\
         \hline        
         Kaon detection efficiency & $ 11\%$ \\
         \hline
         Electron  detection efficiency & $< 2\%$ \\
         \hline
         Radiative corrections & $2.1\%$ \\
         \hline
         Vector meson production & $1\%$ \\
         \hline
         Luminosity determination & $\lesssim 3\%$ \\        
         \hline
         \hline
         Total & $\lesssim 18\%$ \\
         \hline
    \end{tabular}
    \caption{Systematic uncertainty budget of the unpolarized cross section for the produced (final-state) charged kaons.}
    \label{tab:sys_tab_kaon}
\end{table}

In our proposed experiment, there will also be one hour of data collection on a H$_2$ target at both 2.2~GeV and 4.4~GeV beam energies for the purpose of collecting elastic 
scattering data during the future SoLID data acquisition period. This will allow us to reassess the acceptance correction aspect of the current systematic uncertainty analysis, 
by having the SoLID elastic data at our disposal. With high-precision elastic data (from H$_{2}$ reference cell runs) and with possibly improved theoretical models, approximately 
five-ten years from now, we may even have systematic uncertainty re-estimation for the acceptance correction to be much smaller than what is shown in Table~\ref{tab:sys_tab} 
and Table~\ref{tab:sys_tab_kaon}. Our perspective on a better re-evaluation of this component of the total systematic uncertainty is also based on actual detector characteristics: 
large acceptance, high luminosity and kinematic range of the SoLID apparatus, by which it will have better performance in cross-section measurements as compared to BigBite and 
CLAS12 detectors at JLab. In the case of kaons, the large systematic uncertainty of the $13\%$ coincidence acceptance correction arises primarily from the limited analysis currently 
available. To improve this estimate, we plan to incorporate results from the upcoming CLAS12 charged kaon SIDIS analysis \cite{CLAS12Kaon}, which is expected to be published prior 
to the start of SoLID data taking.

Eventually, due to the total systematic uncertainty reduction for both pions and kaons, we will have improved SoLID SIDIS impact results from the MAP ansatz compared to what is shown 
in Fig.~\ref{fig:fig_pika_impact1}
 and Figs.~
\ref{fig:fig_pika_impactC3}, 
\ref{fig:fig_pika_impactC6}. Besides, the scatter plots of average squared transverse momenta in 
Fig.~\ref{fig:fig_scatter_plot1} will be remade as well, with more precise determinations of $\langle P_{\perp}^{2} \rangle_{r}$ and $\langle k_{\perp}^{2} \rangle_{r}$.

\renewcommand{\theequation}{B\arabic{equation}}
\setcounter{equation}{0}
\renewcommand{\thefigure}{B\arabic{figure}}
\renewcommand{\theHfigure}{B\arabic{figure}}
\setcounter{figure}{0}

\section{\label{sec:appB} \\
Azimuthal-angle projections of the SoLID SIDIS unpolarized cross section at low-$Q^2$ region from the general ansatz of Sec.~\ref{sec:phi-dependent}}

By utilizing the simplified TMD framework from Sec.~\ref{sec:phi-dependent}, we can have some insight into azimuthal modulation effects, by adopting the given Gaussian ansatz, 
where intrinsic transverse momentum widths in the unpolarized SIDIS process come from the relative magnitudes of the structure functions $F_{UU}$, $F^{\cos(\phi_h)}_{UU}$, 
and $F^{\cos(2\phi_h)}_{UU}$. 

In this regard, we consider three scenarios that have been used in a data comparison in the analysis of Ref.~\cite{JeffersonLabHallA:2016ctn,Yan:2017}, where two of them give 
good agreement with data in many bins. Namely,
\begin{itemize}
\item[$\bullet$] the model considered in \texttt{Barone2015} \cite{Barone:2015ksa,JeffersonLabHallA:2016ctn} with 
$\langle k_{\perp}^{2} \rangle = 0.037~{\rm (GeV/c)^{2}}$ and $\langle P_{\perp}^{2}\rangle = 0.126 + 0.506 z_{h}^{2}~{\rm (GeV/c)^{2}}$. Within this model, the outgoing 
hadron's transverse momentum is mainly due to the fragmentation process. The $P_{\perp}^{2}$ parametrization gives a good description of the unpolarized SIDIS cross-section 
data from \cite{JeffersonLabHallA:2016ctn} and HERMES data on multiplicities~\cite{Barone:2015ksa}.

\item[$\bullet$] the model considered in \texttt{Bacchetta2011} \cite{Bacchetta:2011gx} with $\langle k_{\perp}^{2}\rangle = 0.14~{\rm (GeV/c)^{2}}$ and 
$\langle P_{\perp}^{2}\rangle = 0.42\,z_{h}^{0.54} (1 - z_{h})^{0.37}~{\rm (GeV/c)^{2}}$. This model too allows for a good description of the unpolarized 
SIDIS cross-section and HERMES multiplicity data.
\end{itemize}

\begin{figure}[h!]
\hskip -0.25truecm
\includegraphics[width=1.015\linewidth]{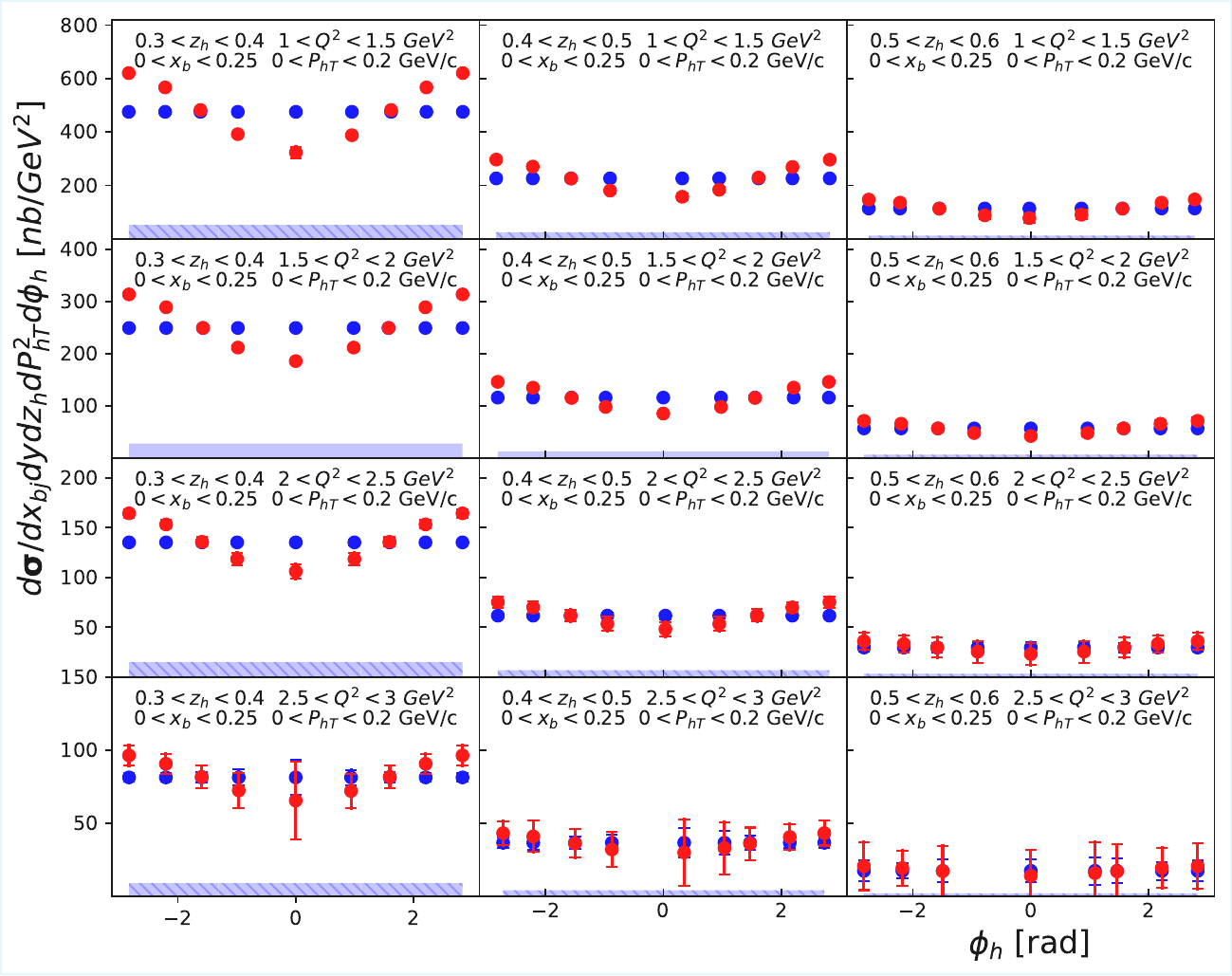}
\vskip 0.0truecm
\caption{Unpolarized cross section for $\pi^+$ production at 11~GeV in selected kinematic bins: 
$x_{bj}=[0,0.25]$, ~$P_{hT}=[0,0.2]$~GeV/c, ~$z_h=[0.3,0.6]$, ~$Q^2=[1,3]$~GeV$^2$. The blue point pseudo-data depict the cross section without the azimuthal modulations, 
and the cross section with the azimuthal modulations are the red points. The bottom band in each plot describes the total systematic uncertainty from Table~\ref{tab:sys_tab}, 
applied to the blue points only.}
\label{fig:fig_xs_pip1}
\end{figure}
\begin{figure}[h!]
\hskip -0.25truecm
\includegraphics[width=1.015\linewidth]{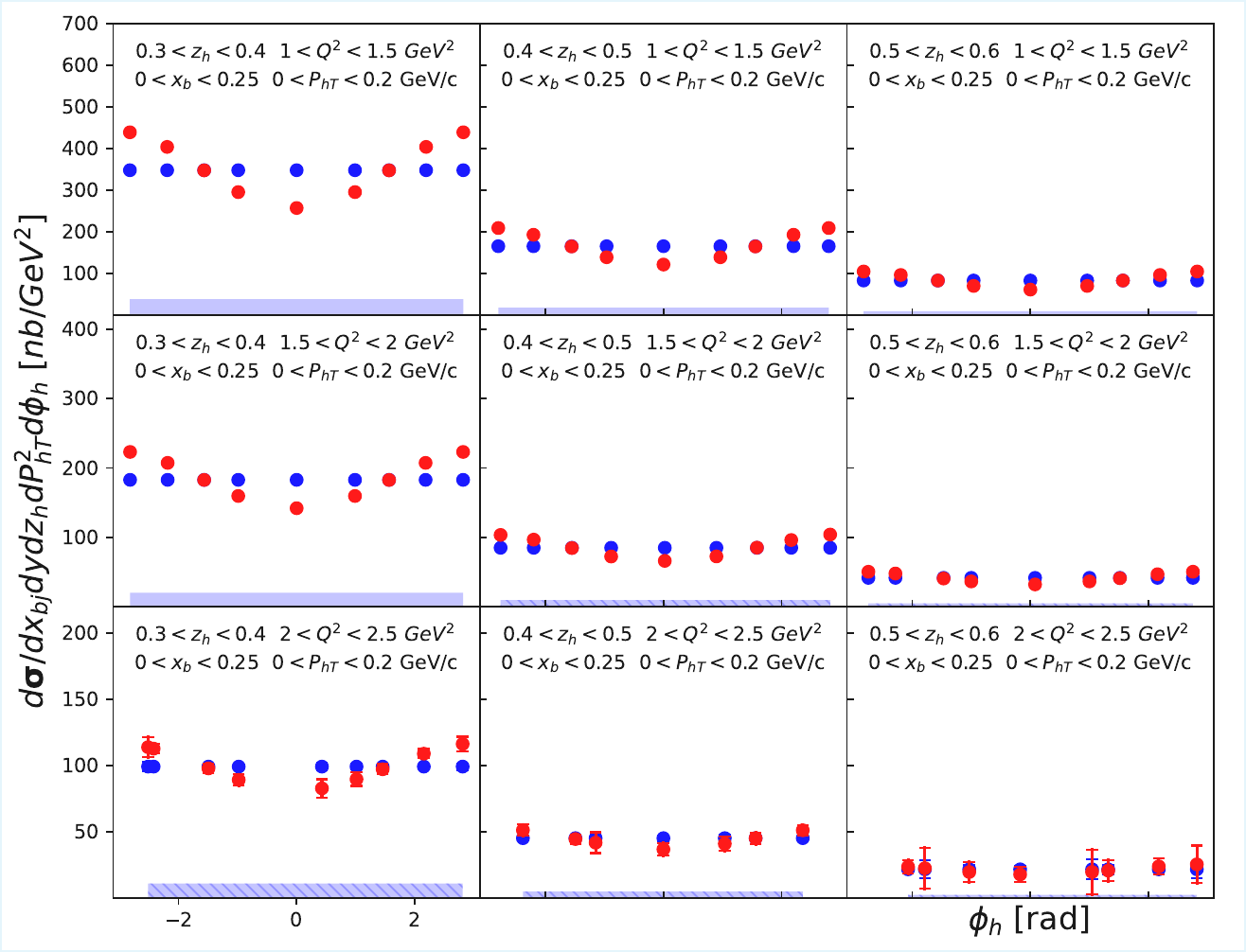}
\vskip 0.0truecm
\caption{Unpolarized cross section for $\pi^+$ production at 8.8~GeV in selected kinematic bins: $x_{bj}=[0,0.25]$, ~$P_{hT}=[0,0.2]$~GeV/c, ~$z_h=[0.3,0.6]$, ~$Q^2=[1,2.5]$~GeV$^2$. 
The rest of the explanation is the same as in Fig.~\ref{fig:fig_xs_pip1}.
}
\label{fig:fig_xs_pip1_8GeV}
\end{figure}
\begin{figure}[h!]
\hskip -0.25truecm
\includegraphics[width=1.015\linewidth]{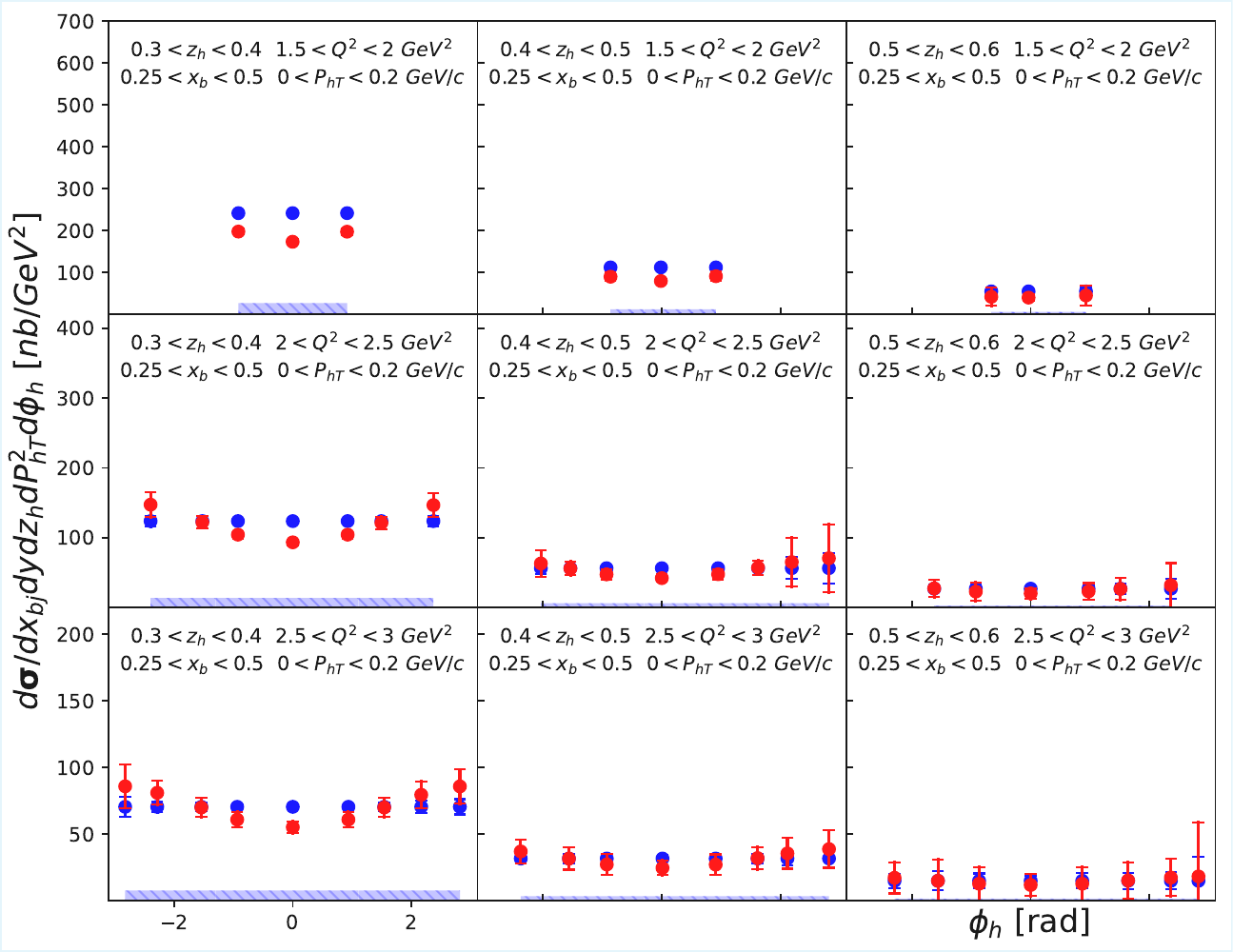}
\vskip 0.0truecm
\caption{Unpolarized cross section for $\pi^+$ production at 11~GeV in selected kinematic bins: $x_{bj}=[0.25,0.5]$, ~$P_{hT}=[0,0.2]$~GeV/c, ~$z_h=[0.3,0.6]$, ~$Q^2=[1.5,3]$~GeV$^2$. 
The rest of the explanation is the same as in Fig.~\ref{fig:fig_xs_pip1}.
}
\label{fig:fig_xs_pip3}
\end{figure}
\begin{figure}[h!]
\hskip -0.25truecm
\includegraphics[width=1.015\linewidth]{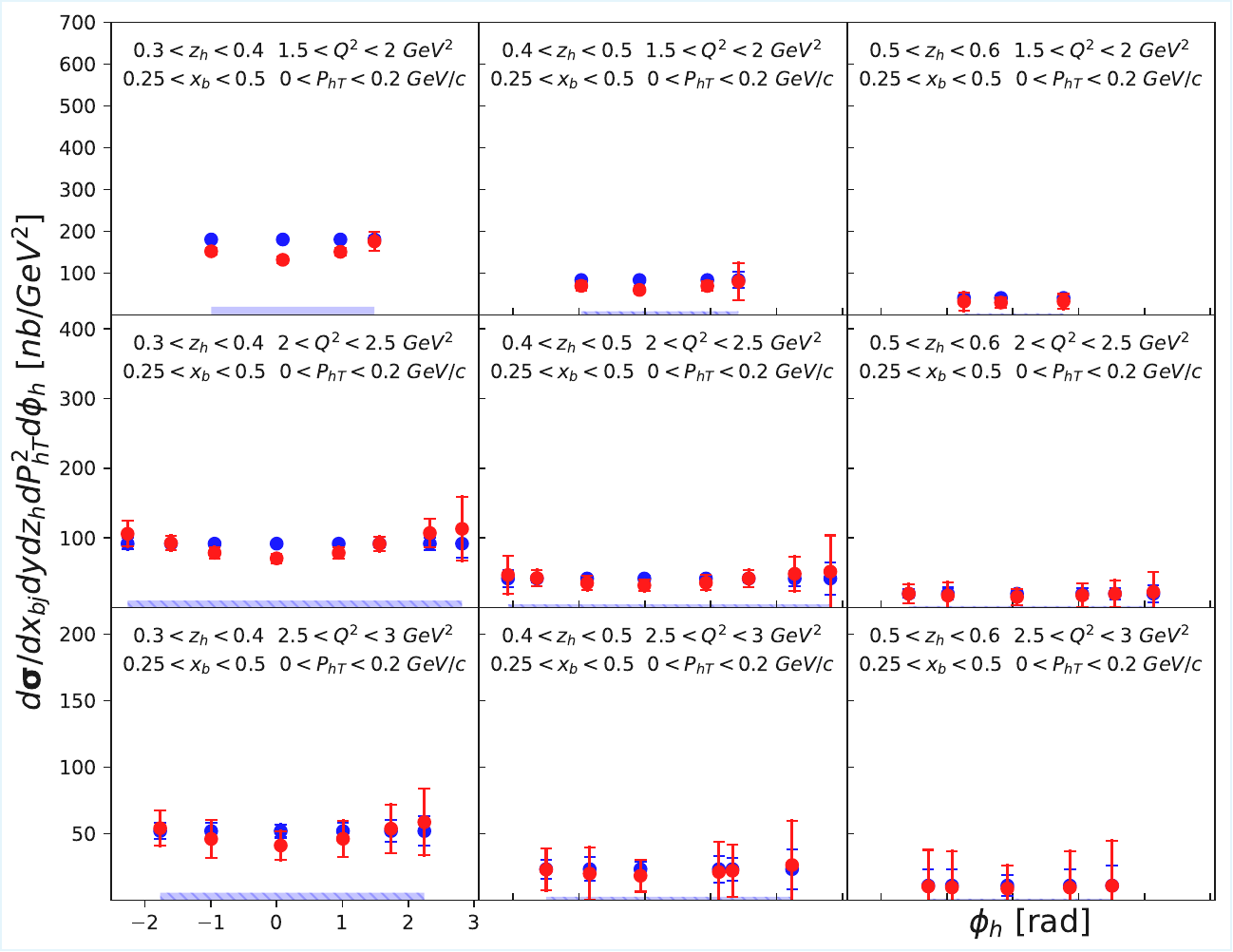}
\vskip 0.0truecm
\caption{Unpolarized cross section for $\pi^+$ production at 8.8~GeV in selected kinematic bins: $x_{bj}=[0.25,0.5]$, ~$P_{hT}=[0,0.2]$~GeV/c, ~$z_h=[0.3,0.6]$, ~$Q^2=[1.5,3]$~GeV$^2$. 
The rest of the explanation is the same as in Fig.~\ref{fig:fig_xs_pip1}.
}
\label{fig:fig_xs_pip3_8GeV}
\end{figure}

In order to generate the unpolarized cross-section pseudo-data in the simplified TMD framework, we adopt another approach that is very close to \texttt{Barone2015} model but 
without any $z_{h}$ dependence in $\langle P_{\perp}^2\rangle$. We call it the \texttt{Default model} taking $\langle k_{\perp}^{2}\rangle = 0.604~{\rm (GeV/c)^{2}}$ and 
$\langle P_{\perp}^{2}\rangle = 0.114~{\rm (GeV/c)^{2}}$ (see Eq.~(\ref{eq:eqn_kT_PT})). The \texttt{Default}, \texttt{Barone2015} and \texttt{Bacchetta2011} models 
are expected to render similar physics representations.

We show here some of our results at low $Q^2$, see Figs.~\ref{fig:fig_xs_pip1}, \ref{fig:fig_xs_pip1_8GeV}, \ref{fig:fig_xs_pip3}, \ref{fig:fig_xs_pip3_8GeV} (see also 
Appendices~A and~B of Ref.~\cite{RG_E12-10-006_12-11-007} for more figures)\footnote{For the unpolarized cross-section expression (with the \texttt{Default model}) see 
Eq.~(\ref{eq:eqn_FUU}), along with Eqs.~(\ref{eq:eqn_FUU_TMD_final})-(\ref{eq:eqn_FUU_PT_final}), Eq.~(\ref{eq:eqn_FUU_cos}) and Eq.~(\ref{eq:eqn_FUU_cos2}).}. In all 
these multi-binned figures, the systematic and statistical uncertainties are the SoLID statistical uncertainties. All details are given in the figure captions.

\subsection*{{}\\
Azimuthal-angle projections of the SoLID SIDIS unpolarized cross section at high-$Q^2$ region from the general ansatz of Sec.~\ref{sec:phi-dependent}}

\begin{figure}[h!]
\hskip -0.25truecm
\includegraphics[width=1.015\linewidth]{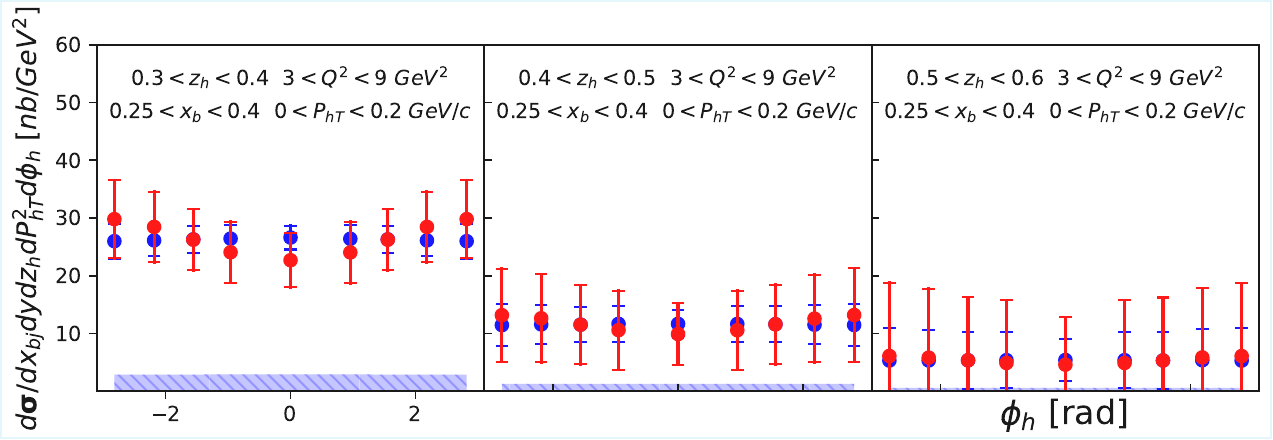}
\vskip 0.0truecm
\caption{
Unpolarized cross section for $\pi^+$ production at 11~GeV in selected kinematic bins: $x_{bj}=[0.25,0.4]$, ~$P_{hT}=[0,0.2]$~GeV/c, ~$z_h=[0.3,0.6]$, ~$Q^2=[3,9]$~GeV$^2$. 
}
\label{fig:fig_xs_highQ2}
\end{figure}

One example at high $Q^2$ is in turn shown in Fig.~\ref{fig:fig_xs_highQ2}. 
Similarly, the bottom band in each plot describes the total systematic uncertainty from Table~\ref{tab:sys_tab}, applied to the blue points only.
 
\newpage
\renewcommand{\theequation}{C\arabic{equation}}
\setcounter{equation}{0}
\renewcommand{\thefigure}{C\arabic{figure}}
\renewcommand{\theHfigure}{C\arabic{figure}}
\setcounter{figure}{0}

\section{\label{sec:appC} \\
Additional MAP results for Sec.~\ref{sec:unpol_res2}}

Here we present the impact study results for the TMD PDFs, 
\begin{figure}[h!]
\hskip -0.25truecm
\includegraphics[width=0.975\linewidth]{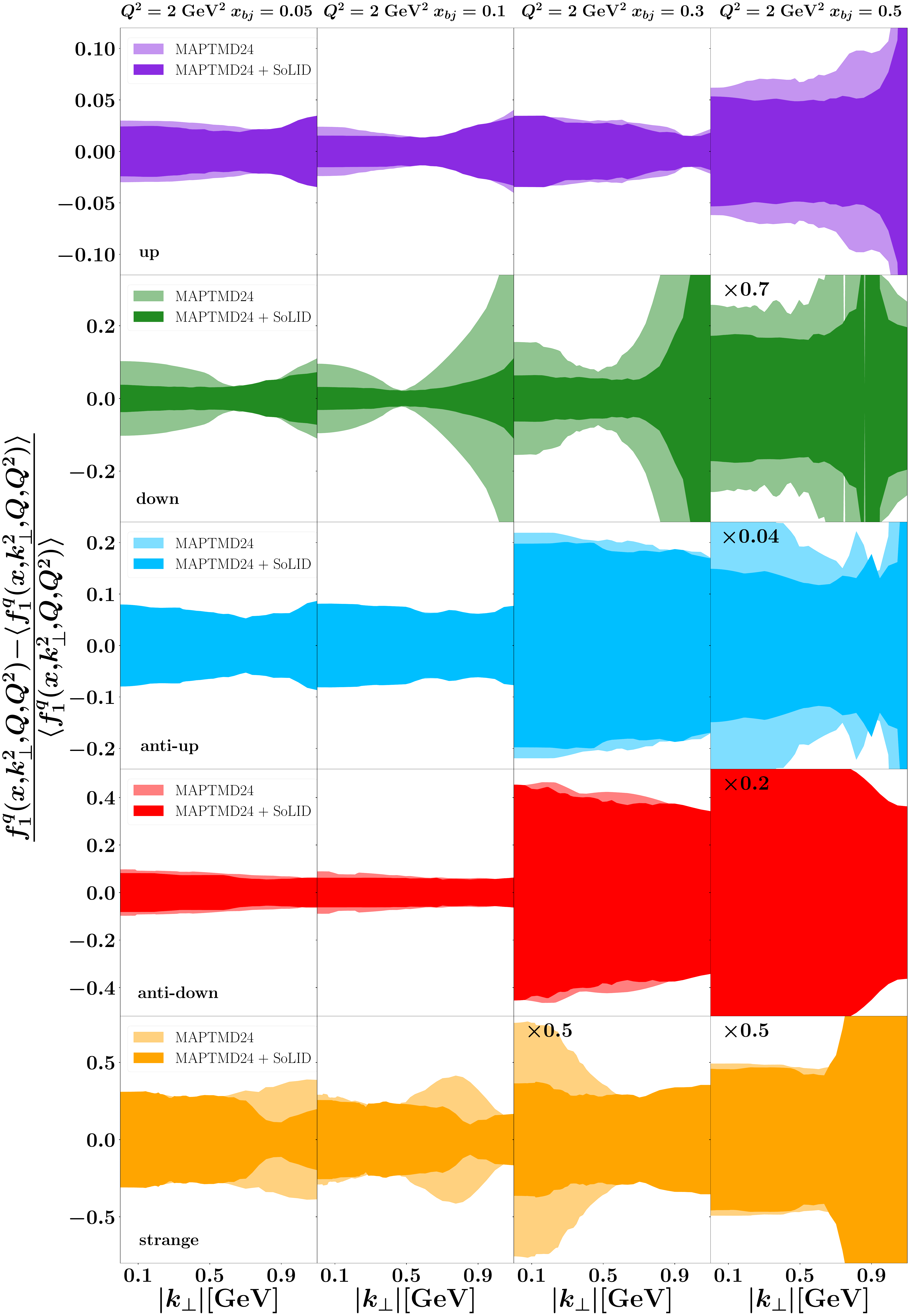}
\vskip 0.00truecm
\caption{The \texttt{MAPTMD24} and impact \texttt{MAPTMD24+SoLID} results on unpolarized TMDs for final-state pions in the bins 
of $x_{{\!}_{bj}} = 0.05$, $x_{{\!}_{bj}} = 0.1$, $x_{{\!}_{bj}} = 0.3$, $x_{{\!}_{bj}} = 0.5$ at $Q^{2} = 2~{\rm GeV^{2}}$ 
as a function of $|\bm{k}_{\perp}|$.}
\label{fig:fig_pika_impactC3}
\end{figure}
Eq.~(\ref{eq:eqn_impact_eq}), by considering pion and kaon production separately, still at fixed $Q^2=2$~GeV$^2$, for different values of 
\xbj = 0.05, 0.1, 0.3 and 0.5, as a function of $|\bm{k}_\perp|$. 
\begin{figure}[h!]
\hskip -0.25truecm
\includegraphics[width=0.975\linewidth]{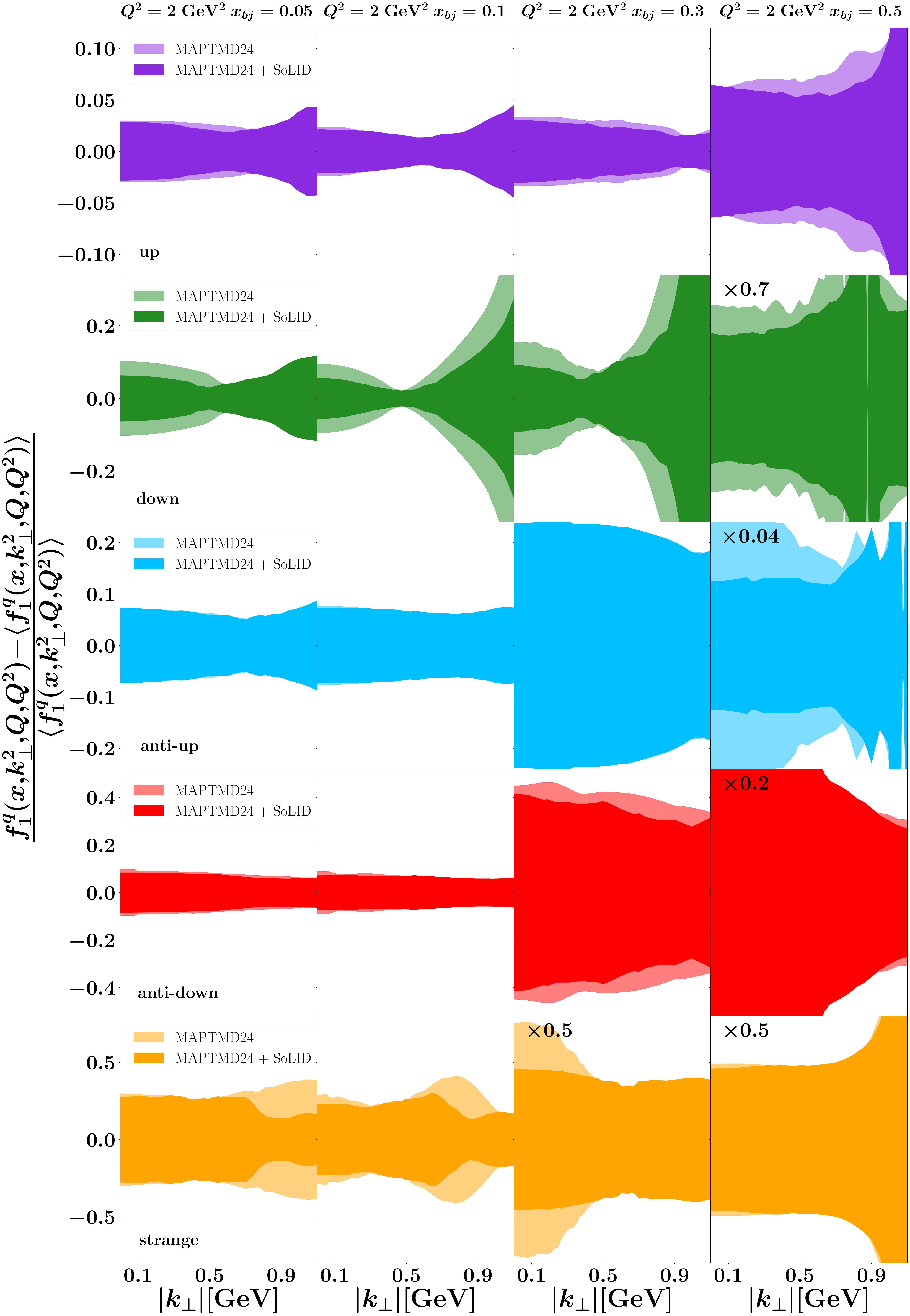}
\vskip 0.00truecm
\caption{The \texttt{MAPTMD24} and impact \texttt{MAPTMD24+SoLID} results on unpolarized TMDs for final-state kaons in the bins 
of $x_{{\!}_{bj}} = 0.05$, $x_{{\!}_{bj}} = 0.1$, $x_{{\!}_{bj}} = 0.3$, $x_{{\!}_{bj}} = 0.5$ at $Q^{2} = 2~{\rm GeV^{2}}$ 
as a function of $|\bm{k}_{\perp}|$.}
\vspace*{2.1in}
\label{fig:fig_pika_impactC6}
\end{figure}

\newpage

\end{document}